\documentclass[acmsmall,screen,nonacm]{acmart}

\acmBooktitle{}

\usepackage{cleveref}
\usepackage{mathpartir}
\usepackage{subcaption}
\usepackage{amsmath}
\usepackage{hyperref}
\usepackage{fancyvrb}
\usepackage{booktabs}

\newlength{\codeindent}
\newcommand{\Infer}[3]{\ensuremath{\inferrule*[right={#1}]{#2}{#3}}}
\newcommand{\ttt}[1]{\texttt{#1}}
\newcommand{\tsc}[1]{\textsc{#1}}
\newcommand{\fin}[1]{\ttt{Fin}~{#1}}
\newcommand{\defunc}{~\rightsquigarrow~}

\newcommand{\letx}[3]{\ttt{let}~#1~\ttt{=}~#2~\ttt{in}~#3}
\newcommand{\lam}[3]{\ttt{\textbackslash} #1\!:\!#2 .~ #3}
\newcommand{\lamnoann}[2]{\ttt{\textbackslash} #1.~ #2}
\newcommand{\view}[3]{\ttt{view}~#1\!:\!#2 .~ #3}
\newcommand{\forexpr}[3]{\ttt{for}~#1\!:\!#2 .~ #3}
\newcommand{\case}[3]{\ttt{case}~#1~\ttt{of}~
                         \ttt{Left  $x$} \rightarrow #2;~
                         \ttt{Right $x$} \rightarrow #3}
\newcommand{\casebranch}[3]{\ttt{case}~#1~\ttt{of}~#2;~#3}
\newcommand{\subst}[3]{[#1\!\mapsto\!#2]#3}
\newcommand{\ann}{\!:\!}

\newcommand{\R}{\mathbb{R}}

\newcommand{\codelinesize}{\normalsize}

\newcommand{\diffenv}{\Delta}
\newcommand{\linruleop}[2]{\mathcal{D}_{#1}[ #2 ] }
\newcommand{\linrule}[2]{\linruleop{#1}{#2} \rightsquigarrow }
\newcommand{\linreifyop}[3]{#1\vdash \mathcal{L}_{#2}[ #3 ] }
\newcommand{\linreify}[3]{\linreifyop{#1}{#2}{#3} \rightsquigarrow }
\newcommand{\linresult}[3]{#1,~#2,~#3}
\newcommand{\zeroat}[1]{\ttt{Zero}[#1]}
\newcommand{\tangentof}[1]{\ttt{Tan}[#1]}

\newcommand{\transruleop}[3]{\mathcal{T}_{#1}[#2,~#3] }
\newcommand{\transrule}[3]{\transruleop{#1}{#2}{#3} \rightsquigarrow }
\newcommand{\transenv}{\Omega}
\newcommand{\action}[4]{\ttt{\textbackslash}#1\ann\ttt{Type}~#2\ann#3\ttt{.}~#4}

\newif\ifextended
\extendedtrue

\newcommand{\ExcessDetail}[1]{}

\begin{document}

% \title{Dex: A functional, stateful, and parallel numerical language}
\title[Getting to the Point.]{Getting to the Point. \\
\small Index Sets and Parallelism-Preserving Autodiff for Pointful Array Programming}

\author{Adam Paszke}
\email{apaszke@google.com}
\affiliation{%
  \institution{Google Research}
  \country{Poland}
}

\author{Daniel Johnson}
\email{ddjohnson@google.com}
\affiliation{%
  \institution{Google Research}
  \country{Canada}
}

\author{David Duvenaud}
\email{duvenaud@cs.toronto.edu}
\affiliation{%
  \institution{University of Toronto}
  \country{Canada}
}

\author{Dimitrios Vytiniotis}
\email{dvytin@google.com}
\affiliation{%
  \institution{DeepMind}
  \country{United Kingdom}
}

\author{Alexey Radul}
\email{axch@google.com}
\affiliation{%
  \institution{Google Research}
  \country{USA}
}

\author{Matthew Johnson}
\email{mattjj@google.com}
\affiliation{%
  \institution{Google Research}
  \country{USA}
}

\author{Jonathan Ragan-Kelley}
\email{jrk@mit.edu}
\affiliation{%
  \institution{Massachusetts Institute of Technology}
  \country{USA}
}

\author{Dougal Maclaurin}
\email{dougalm@google.com}
\affiliation{%
  \institution{Google Research}
  \country{USA}
}

\renewcommand{\shortauthors}{Paszke, Johnson, Duvenaud, Vytiniotis, Radul, Johnson, Ragan-Kelley, Maclaurin}

\begin{abstract}
We present a novel programming language design that attempts to combine the clarity and safety of high-level functional languages with the efficiency and parallelism of low-level numerical languages.
We treat arrays as eagerly-memoized functions on typed index sets, allowing abstract function manipulations, such as currying, to work on arrays.
In contrast to composing primitive bulk-array operations, we argue for an explicit nested indexing style that mirrors application of functions to arguments.
We also introduce a fine-grained typed effects system which affords concise and automatically-parallelized in-place updates.  Specifically, an associative accumulation effect allows reverse-mode automatic differentiation of in-place updates in a way that preserves parallelism.
Empirically, we benchmark against the Futhark array programming language, and demonstrate that aggressive inlining and type-driven compilation allows array programs to be written in an expressive, ``pointful'' style with little performance penalty.
\end{abstract}
% Empirically, we benchmark against the Futhark array programming language, and demonstrate that aggressive defunctionalization allows extensive use of typeclasses and higher-order functions with little performance penalty.




\maketitle

\section{Introduction}
\label{sec:introduction}

Recent years have seen a dramatic rise in the popularity of the array  programming model.
The model was introduced in APL \cite{iverson1962apl}, widely popularized by MATLAB, and eventually made its way into Python. There are now tens of libraries centered around $n$-dimensional arrays (nd-arrays), each with its own unique strengths.
In fact, this model is now so important that many hardware vendors often market their products by evaluating various nd-array workloads, and publish their own libraries of array kernels (e.g., NVIDIA's cuBLAS) or compilers for array programs (e.g., Google's XLA).
We theorize that the nd-array model's recent success is due to a convergence of three factors: its reasonable \emph{expressiveness} for concise, high-level specification of (relatively) large workloads; the \emph{abundant parallelism} available within the individual nd-array operations; and the \emph{preservation of parallelism under automatic differentiation}.

Preserving parallelism under automatic differentiation is a particularly important advantage.
The trouble is that the most popular automatic differentiation algorithm (the so-called reverse-mode) inverts all data dependencies in the original program.
This unfortunately turns perfectly parallel loops that read a datum repeatedly into sequential loops that have to write to the cotangent for that datum many times.
Most modern nd-array libraries circumvent this problem by operating at a higher level of abstraction: instead of differentiating through the individual array accesses used to implement each operation (e.g., scalar multiply-adds in matrix multiplication), the high-level operations themselves are replaced with new high-level operations that compute the appropriate cotangents in bulk.%
\footnote{Such transformation rules can either be written by hand for each operation \cite{jax2018github}, or derived using special-purpose techniques \citep{huckelheim2019automatic}.}
The result is that the reverse-AD transform for a typical nd-array program produces another typical nd-array program, namely a sequential composition of parallel blocks.

Forcing the original program to be written at a higher level of abstraction, however, has significant downsides.
The first major flaw, in our view, is \emph{insufficient expressiveness}: while composing a program out of array operations leads to wonderfully fast and concise code when the available array operations cover one's needs, one faces a steep complexity and performance cliff---dropping down to manually writing, optimizing, and differentiating new operations---as soon as they do not.
%\DAVE{This paragraph seems worth spending time on, since it's clarifying the scope where Dex should be a big improvement.}
Examples of algorithms that are difficult to express efficiently in this style include differential equation solvers, natural language processing algorithms, weather and physics simulations, and workloads that need non-linear stencil operations.
In these domains, inner loops are still regularly written in loosely-typed, low-level Fortran or CUDA.
%\JRK{Citations?}

The model's second major flaw is \emph{insufficient clarity}: for their code size, nd-array programs are notoriously difficult to read, reason about, and get right.
Practitioners constantly complain about ``shape errors,'' which are really mis-specification of the intended flow of data.
Indeed, when an entire multidimensional array is referred to by a single variable name, its internal semantics (for instance, which dimensions are indexing what) are missing from the program---and have to be continuously reconstructed by the programmer, which is both tedious and error-prone. %engendering tedium and propensity for error.
Logic errors are especially costly in two of the classic applications of array programming---stochastic simulation and learning algorithms---because tests are less effective in these domains.

% \JRK{Do we need to anonymize Dex? Should we at least replace it with a macro, so we can decide on this later?}
This paper introduces Dex, a new array programming language oriented around safe, efficient, ergonomic, and differentiable typed indexing.
The contributions are as follows:
\begin{itemize}
  \item Typed indexing constitutes a novel, concise yet flexible notation for expressing array programs (\Cref{sec:typed-indexing}).
  \item Typing array index sets creates an analogy between
  arrays and functions, which in turn leads to a fruitful convergence of array and functional
  programming (\Cref{sec:function-array-duality}).
%   \DV{Technically the F-smooth approach is very similar, so I am going back and forth about the
%   "novel" part. The novelty is that this approach is combined with efficient AD that preserves
%   parallelism thanks to runAccum, which was clearly missing in F-smooth}
%   \DV{is there an argument that the expressive types allow you also to write code that's faster,
%   cause you have to tag less at runtime. Perhaps if you introduce union types or dynamic programming
%   examples.? That would be an interesting boost to the argument.}
  \item We introduce an effect for associative accumulation in Dex's effect system. The \texttt{Accum} effect hits a sweet spot of expressiveness and parallelizability (\Cref{sec:effects}). % \JRK{Sweet for what? AD? Numerics in general? Feels fuzzy right now.}
  \item We improve upon the defunctionalization methods presented in \citet{hovgaard2018high}, in a way that allows functions to be treated as first-class objects and can deal with arrays of functions or returning functions from conditionals (\Cref{sec:defunctionalization}).
  % \DANIEL{Reminder that we still say ``defunctionalization'' here. Is this a correct usage?}
  % \AXCH{The title of the reference says "Defunctionalization", so I would say "yes".}
    % \DV{Is that a novel contribution, or
%     something that we know how to do in the functional programming community, such as typed closure
%     conversion and defunctionalization with higher-order functions? If it's not a novel thing perhaps
%     you should reword to: Dex is also novel amongst array languages in that it includes an expresssive
%     defunctionalization pass (along the lines of \cite{xyz}), that can .... I.e. qualify it a bit "novel amongst array languages" to not give the wrong impressions to readers.}
  \item We implement efficient, parallelizable automatic differentiation with full coverage of Dex, including indexing (\Cref{sec:ad}).
  \item We articulate several lessons we learned from treating AD as a first-class concern while designing the Dex language (\Cref{sec:ad-lessons}).
  \item We describe our strategy for achieving competitive performance (\Cref{sec:performance}).  The main ingredients are:
  \begin{itemize}
    \item Work-efficient automatic parallelization (preserved through AD) via the \ttt{Accum} effect (\Cref{sec:auto-parallel}).
    \item Straightforward loop fusion via inlining \ttt{for} expressions (\Cref{sec:fusion}).
    \item Index-set-type directed compilation to elide bounds checks and make layout decisions (\Cref{sec:type-directed-compilation}).
  \end{itemize}
\end{itemize}

\section{Dex by Example}
\label{sec:expressiveness}

Dex is a strict functional language for index-oriented\footnote{``Pointful'', as opposed to point-free.} array programming defined by a confluence of several features:
\begin{itemize}
    \item Typed array dimensions and indices (\Cref{sec:typed-indexing}),
    \item A parallelism-friendly effect system (\Cref{sec:effects}), and
    \item Work- and parallelism-preserving automatic differentiation (\Cref{sec:ad}).
\end{itemize}
We start with building intuition and giving examples, especially in \Cref{sec:mandelbrot}.
A formal description of the language can be found in~\Cref{sec:formal-language}.

\subsection{Arrays with Typed Indexing}
\label{sec:typed-indexing}

% The Dex array type is written \texttt{n=>a}.
% Here \texttt{n} is the type of the array's indices, which we refer to as the array's \emph{index set}; while \texttt{a} as usual is the type of the array's elements.
Dex defines two syntactic forms for manipulating arrays.
The first form is used to construct arrays:
$$\ttt{array = for i:n. expr}$$
The \texttt{for} expression is similar to the \texttt{build} function of $\tilde{F}$ \cite{shaikhha2019fsmooth}.
The result of such a \texttt{for} expression is an array constructed by repeatedly evaluating the body, \texttt{expr}, with \texttt{i} bound to each consecutive member of the \emph{index set} \texttt{n}.
The array will thus have as many elements as \texttt{n} has members.
As we will see in the next few examples, \ttt{n} can usually be omitted, because Dex infers it from the body.

% A common simple form of index sets is \emph{Fin k} where $k$ is value of natural number type, denoting
% the values between $0$ and $k-1$. For instance here's an expression that builds the array $[1,2,...,32]$: $$\mathtt{for\ i:Finite(32). i+1}$$.
% While index sets are just ordinary Dex types, not every type in the language can be used as an index set.
% We provide an extensive discussion in \cref{sec:index-sets}.

The second form is array indexing, written using the \texttt{(.)} operator:
$$ \ttt{element = array.j} $$
The index \texttt{j} must be a member of the array's index set \texttt{n}, and \texttt{array.j} extracts the \texttt{j}th element.
These two operations are the fundamental building blocks of all Dex programs.

The magic sauce is the type of arrays.  We type an array as
$$\texttt{array : [IndexSet n] n=>a},$$
which exposes both the index set \ttt{n} and the element type \ttt{a} to the type system.%
\footnote{The type system becomes value-dependent to accommodate index sets; more in \Cref{sec:formal-type-system}.}
Array elements can be any type, including functions and other arrays, but array indices have to obey an \ttt{IndexSet} class constraint.  More on this in \Cref{sec:index-sets}.

As our first example, consider transposing a matrix \texttt{matrix}:
\begin{Verbatim}[xleftmargin=\codeindent,fontsize=\codelinesize]
transposed = for i j. matrix.j.i
\end{Verbatim}
This expression loops over two indices (one for each dimension of \texttt{matrix}) in the order \texttt{i} then \texttt{j}, but then applies them in reverse.
The same pattern can be easily adapted to reorder any number of dimensions of an $n$-dimensional array.
Notice that we didn't have to give iteration bounds in the \texttt{for} expression---Dex's type system infers them from the type of \texttt{matrix}.

We can just as easily multiply two matrices \texttt{x} and \texttt{y}.
\begin{Verbatim}[xleftmargin=\codeindent,fontsize=\codelinesize]
x_times_y = for i j. sum (for k. x.i.k * y.k.j)
\end{Verbatim}
In this example, all the \texttt{for}-bound variables are again used in array indexing, meaning that their ranges can be inferred from the types of \texttt{x} and \texttt{y}.
Explicit indexing makes it straightforward to work out the semantics of this, even if one is not already familiar with it.
Reading the expression inside out:
\begin{enumerate}
  \item We first construct element-wise products of vectors selected from the first dimension of the first input (rows) and second dimension of the second input (columns): \texttt{for k. x.i.k * y.k.j}.
  \item Then, we take the sum of these products by calling the function \texttt{sum} (which we return
  to in~\Cref{sec:effects}).
  \item This computation is repeated for all valid indices \texttt{i} and \texttt{j} (\texttt{for i j.}). Since the result of \texttt{sum} is a scalar, the result \verb|x_times_y| of the repeated computation is a two-dimensional array (i.e., a matrix).
\end{enumerate}

Note that we used the same index, \texttt{k}, for the second dimension of \texttt{x} as for the first dimension of \texttt{y}.
Dex's type system will therefore statically check those array dimensions for equality.
Typed index sets make a whole class of shape errors easier to detect and repair in Dex.

\subsection{Complete Example}
\label{sec:mandelbrot}

\begin{figure}
\begin{Verbatim}[xleftmargin=2cm,fontsize=\small]
def update (c:Complex) (z:Complex) : Complex = c + (z * z)

def inBounds (z:Complex) : Bool = complex_abs z < 2.0

def escapeTime (c:Complex) : Int =
  fst $ yieldState (0, zero) \(n, z).
    for i:(Fin 1000).
      z := update c $ get z
      n := (get n) + (BToF $ inBounds $ get z)

xs = linspace (Fin 300) (-2.0) 1.0
ys = linspace (Fin 200) (-1.0) 1.0

mandelbrot = for j i. escapeTime (MkComplex xs.i ys.j)
\end{Verbatim}
\caption{Computing the Mandelbrot set in Dex.}
\label{fig:mandelbrot}
\end{figure}

Other than typed indexing and the effect system we discuss in \Cref{sec:effects}, the core of Dex looks like a strict functional programming language in the ML \citep{milner1997ml} syntactic family.
\Cref{fig:mandelbrot} shows a complete example Dex program for computing the Mandelbrot set.
A few points to pay attention to when parsing this:
\begin{itemize}
\item The colon \text{:} is Dex for ``has type'', and introduces the mostly-optional type annotations for variable bindings and function return values.
\item The back-slash \verb|\| is a lambda function expression: \verb|\n z. body| is a function that accepts arguments \texttt{n} and \texttt{z} (in this case, references in the \texttt{State} effect).
\item Note the syntactic overloading of the dot (\texttt{.}). It serves as a decimal point, as the array indexing operator, and as the delimiter between the binders and body of \texttt{for} and \verb|\| expressions.
\item The \texttt{(Fin 1000)} is an index set representing 1000 elements.  We will cover \texttt{Fin} when we discuss index sets in \Cref{sec:index-sets}; in this program, it indicates that our loop should iterate 1000 times, and specifies the number of elements in the grid returned by \texttt{linspace}.
\end{itemize}

The \ttt{yieldState} function is a convenience wrapper which returns the final state after running a stateful action (using the primitive \ttt{runState}).
We now proceed to a more in-depth discussion of Dex's effect system.

\subsection{Effects}
\label{sec:effects}

Pointwise transforms (maps)---the fundamental building block of array programs---are easy to write with Dex's \texttt{for} loops.
But they are far from being sufficient for an expressive modern array language.
Even the simple matrix multiplication example required a \texttt{sum} function, which cannot be written as a pure \texttt{for} expression because it must combine many different elements of the input array while iterating across the index set.
%because there is no way to make the different elements of the input array meet to be summed.

We thus need a means for controlled communication across different iterations of a \texttt{for}.
To that end, we extend Dex with an \emph{effect system}.
The type of every expression is extended with the set of effects its evaluation will induce.
We give the flavor of Dex's effects by discussing two different ways to spell \texttt{sum}.

\subsubsection{Summing as state}
\label{sec:sum-via-state}
One of the supported effects is \texttt{State}, allowing arbitrary reads and updates to a shared state, which is sufficient to make \texttt{sum} expressible in Dex:
\begin{Verbatim}[xleftmargin=\codeindent,fontsize=\small]
sum = \x:n=>Float.
  (_, total) = runState 0.0 \ref.
    for i.
      ref := (get ref) + x.i
  total
\end{Verbatim}
Let us dissect this snippet line-by-line.
We begin with a lambda definition that binds the vector to be reduced to the variable \texttt{x}, whose type we annotate as \ttt{n=>Float}.
% This lets \ttt{sum} reduce over an arbitrary index set \ttt{n}.
% Since \ttt{n} is not specified, \ttt{sum} can reduce over an arbitrary index set with an arbitrary size.
% Next, while we do use \texttt{State} in this implementation of \texttt{sum}, the effect should only be visible internally and the function itself should be pure (i.e., have no side-effects visible from the outside).
% For that purpose, we use the \texttt{runState} combinator to discharge the effect.
Here \ttt{n} denotes an arbitrary index set, since we want \ttt{sum} to sum over arrays of any size.
Next, we apply the \texttt{runState} combinator, which allows us to locally execute a stateful action within the pure function \texttt{sum}.
\texttt{runState} takes the initial value of type \ttt{s} for the state along with a stateful function, and applies the function once to an appropriately-initialized \emph{mutable reference} of type \ttt{Ref h s}.%
\footnote{The state has to be something we can allocate a mutable buffer for, which we model with the \ttt{Data} class constraint.}
\begin{Verbatim}[xleftmargin=\codeindent,fontsize=\codelinesize]
runState : [Data s] s -> (h:Type ?-> Ref h s -> {State h} a) -> (a, s)
\end{Verbatim}
Once we have the reference \ttt{ref} in scope, we begin looping over indices that are valid for the vector \texttt{x}.
At each step we retrieve the current value of the state using the \texttt{get} function, add to it the \texttt{i}-th value in \texttt{x}, and update the state with the result.
\begin{Verbatim}[xleftmargin=\codeindent,fontsize=\small]
get  : Ref h a -> {State h} a
(:=) : Ref h a -> a -> {State h} Unit
\end{Verbatim}
Finally, \texttt{runState} returns a pair containing the result of its body and the value of the reference once the body has been evaluated.
Since the body does not have any meaningful results (it is of type \texttt{n=>Unit}---an array full of unit values), we skip the first component and return the sum.

The mysterious \texttt{h} parameter appearing in the \texttt{Ref} type implements the classic rank-2 polymorphism trick used for Haskell's ST monad \cite{launchbury1994st}.
The user function given to \texttt{runState} must accept an argument of type \texttt{Ref h s} for an arbitrary local type parameter \texttt{h}, which guarantees that the reference cannot escape the block in which it is valid.
The \texttt{?->} function arrow in Dex indicates that \texttt{h} is an \emph{implicit argument} which can be inferred from context and thus does not have to be written out by the user.

\subsubsection{Summing as accumulation}
\label{sec:sum-via-accum}

Unfortunately, expressing \texttt{sum} with a \ttt{State} effect as above is a mixed bag.
While it does compute the correct sum, it sacrifices parallelism in the process.
Whenever the body of a \texttt{for} expression induces a \texttt{State} effect, it is possible that each loop iteration can have arbitrary dependencies on previous iterations. A sufficiently powerful analysis might let us discover some subset of stateful loops that are parallelizable, but such an analysis is difficult, and in the absence of such an an analysis we are forced to pessimistically evaluate the loop sequentially.
% we pessimistically have to make the loop sequential, because at that point it might have arbitrary dependencies between the different iterations.
% Of course a sufficiently powerful analysis might let us discover some subset of stateful loops that are parallelizable, but it is difficult and the results tend to be unreliable.
% Matrix multiplication is the poster child of modern parallel hardware, with each hardware accelerator having dedicated units for parallel matrix multiplication, and we simply cannot afford to give up the ability to target those.

Fortunately, Dex also includes a more specific \emph{accumulation} effect, which we call \texttt{Accum}.
The accumulation effect is similar to \texttt{State} in that it also exposes mutable references, but \ttt{Accum} imposes two critical restrictions on them.
First, updates in \ttt{Accum} must be \emph{additive contributions} to references with (finite-dimensional) \emph{vector space} types:
\begin{Verbatim}[xleftmargin=\codeindent,fontsize=\codelinesize]
(+=) : [VectorSpace w] Ref h w -> w -> {Accum h} Unit
\end{Verbatim}
Here \texttt{VectorSpace w} denotes a constraint on the (otherwise polymorphic) type \texttt{w}: it must be an array \texttt{n=>b} for some vector space \texttt{b}, a pair \texttt{(b, c)} of vector spaces \texttt{b} and \texttt{c}, or a \texttt{Float}. This constraint is checked by the compiler when typechecking functions that use \texttt{(+=)}.

Second, there is no ``read'' operation in \ttt{Accum}---the value of the accumulator cannot be retrieved until the reference goes out of scope at the end of the \ttt{runAccum} handler (and thus until all writes have completed):
\begin{Verbatim}[xleftmargin=\codeindent,fontsize=\codelinesize]
runAccum : [VectorSpace w] (h:Type ?-> Ref h w -> {Accum h} a) -> (a, w)
\end{Verbatim}

Together, these constraints mean that all updates to a reference in \ttt{Accum} are associative (up to floating-point roundoff), and therefore functions using the \texttt{Accum} effect can be efficiently parallelized.
Specifically, we can partition any sequence of updates into multiple subsequences, compute partial sums for each subsequence, and then finally combine the partial sums.

% Here is \texttt{sum} again, this time using the \texttt{Accum} effect:
Using the \texttt{Accum} effect, we can implement \texttt{sum} in a parallel-friendly way:
\begin{Verbatim}[xleftmargin=\codeindent,fontsize=\small]
sum = \x:n=>Float.
  snd $ runAccum \total.
    for i. total += x.i
\end{Verbatim}

The \ttt{Accum} effect turns out to be flexible enough to cover many numerical algorithms, including automatically generated derivatives, allowing them to be parallelized efficiently (see \Cref{sec:auto-parallel}).
However, since not all algorithms are compatible with the restrictions of \texttt{Accum}, Dex allows the user to choose between parallel-but-write-only \texttt{Accum} and serial-but-flexible \texttt{State}.

Note that in the full surface language \texttt{runAccum} is extended to support arbitrary user-defined \texttt{Monoid} instances in the style of Haskell, which makes the \texttt{Accum} effect essentially isomorphic to the \texttt{Writer} monad.
However, automatic differentiation only needs accumulation on finite-dimensional vector spaces, so for this paper we assume the more restrictive vector space constraint.

\subsubsection{Reference indexing}
\label{sec:reference-slicing}

While the design of our effect system largely follows previous research, with Koka \cite{leijen2014koka} being a huge inspiration, we have extended it a bit further to include \emph{reference indexing}. Specifically, the references used by the {\tt State} and {\tt Accum} effects allow the following {\em pure} operation:
\begin{Verbatim}[xleftmargin=\codeindent,fontsize=\codelinesize]
(!) : Ref h (n=>a) -> n -> Ref h a
\end{Verbatim}
Given a reference to an array and an index, the \texttt{!} operator extracts a reference to an array element.

Many linear-algebra routines often perform operations over, e.g., the rows of a matrix, and being able to ``slice'' references in this way makes it especially convenient to implement such routines, especially since chained reference indexing efficiently takes apart nested arrays.
But it turns out that it is also a crucial component for being able to express parallel patterns such as segmented reductions or histograms (at least in a work-efficient way, see~\ref{sec:parallel-accum}):
\begin{Verbatim}[xleftmargin=\codeindent,fontsize=\small]
histogram = \x:n=>bins.
  snd $ runAccum \hist.
    for i. hist!(x.i) += 1
\end{Verbatim}
Finally, reference indexing is important for the efficiency of reverse-mode automatic differentiation in the presence of indexing. More details can be found in \Cref{sec:ad-of-indexing}.

% \todo[inline]{axch@ asks: Is there an opportunity to parallelize State effects across non-overlapping slices of the same reference?}
% For instance, can each loop iteration (and call to \verb|action|) be executed in parallel in the below combinator, without making any defensive copies?
% \begin{Verbatim}[xleftmargin=\codeindent,fontsize=\small]
% withStateSlices :
%     Ref h1 (n=>s) -> (h2:Type ?-> Ref h2 s -> {State h2} a) -> {State h} n=>a
% withStateSlices = \ table:Ref h1 (n=>s) action.
%   for i.
%     (a, s) = runState (get table!i) action
%     table!i := s
%     a
% \end{Verbatim}
% The \verb|withStateSlices| combinator is almost equivalent to
% \begin{Verbatim}[xleftmargin=\codeindent,fontsize=\small]
% stateOnSlices : n=>s -> (h:Type ?-> Ref h2 s -> {State h} a) -> n=>(a, s)
% stateOnSlices = \ table action.
%   for i. runState table.i action
% \end{Verbatim}
% except that the latter requires its input \verb|table| to be a pure value, which may force a copy for the sake of parallelism if we're already in a context where we have a \verb|Ref h (n=>s)|.

\subsection{Comparisons}
\label{sec:expressiveness-comparisons}

We conclude our informal introduction to Dex by comparing its expressiveness with three other array programming notations.
Our running point of comparison is the matrix multiplication we have already seen, this time written as a Dex function:
\begin{Verbatim}[xleftmargin=\codeindent,fontsize=\small]
def matrix_multiply (x:n=>m=>Float) (y:m=>o=>Float): n=>o=>Float =
  for i j. sum (for k. x.i.k * y.k.j)
\end{Verbatim}

\subsubsection{Expressiveness in comparison to combinator languages}
\label{sec:array-combinator-languages}

Most of the other functional languages for array computing (e.g. Futhark \cite{henriksen2017futhark}, Accelerate \cite{chakravarty2011accelerate}, LIFT \cite{steuwer2017lift}, XLA) take a somewhat different approach than Dex.
Instead of exposing the array building expressions and encouraging explicit indexing, they usually provide a set of array combinators built into the language that are the only way to manipulate arrays.
The combinators are often modeled after higher-order functions common between the functional programming and parallel computing communities, such as \texttt{map}, \texttt{scan}, \texttt{zip}, and \texttt{reduce}.
Since this is the currently prevailing approach to array computing, it is the target of many of our comparisons in this paper, and we will refer to the languages that follow this approach as \emph{array-combinator languages}.

In an array-combinator language, matrix multiplication might be implemented as follows:
\begin{Verbatim}[xleftmargin=\codeindent,fontsize=\small]
combinator_matrix_multiply = \x y.
  yt = transpose y
  dot = \x y. sum (map (uncurry *) (zip x y))
  map (\xr. map (\yc. dot xr yc) yt) x
\end{Verbatim}
While this implementation is quite terse as well, it arguably does not explicate the meaning of matrix multiplication quite as explicitly as the Dex version.
Many simple operations, such as a dot product, have to be expressed using multiple combinators, with explicit zipping when multiple arguments are to be consumed.
Finally, it is worth noting that this style is also available in Dex, and the above is a valid Dex function.

\subsubsection{Expressiveness in comparison to Einstein notation}
\label{sec:einstein-notation}
Another interesting reference point is the Einstein notation, for instance as realized in the domain-specific language for the first arument to the function \texttt{numpy.einsum}.
This function has been hugely successful in recent years and is one of the more popular functions, especially in the field of machine learning.
It generalizes a wide variety of linear algebra operations, including matrix multiplication:
\begin{Verbatim}[xleftmargin=\codeindent,fontsize=\small]
def matrix_multiply(x, y):
  return np.einsum('ik,kj->ij', x, y)
\end{Verbatim}

But, as it turns out, it is also very close to the Dex notation!
Notice how the \texttt{ik} and \texttt{kj} index sequences exactly follow the Dex indexing expressions, while the \texttt{ij} output annotation is exactly the order of \texttt{for} binders.
Hence, Dex can be seen as a sort of generalized Einstein notation, thanks to type inference automatically binding dimension sizes to loop indices.
But Dex is also more flexible, because it is not limited to loops that first multiply and then sum dimensions in the standard semiring of real numbers.
Both of those extensions are similar in spirit to the extensions explored by projects such as Tensor Comprehensions \citep{vasilache2018tensor}.

\subsubsection{Expressiveness in comparison to C}
\label{sec:compare-c++}
The de-facto language for low-level numerical computations is C, which benefits from a large body of research on optimization techniques, especially for numerical programs~\citep{grosser2012polly, pluto}.
A straightfoward matrix multiplication routine in C might look something like this:
\begin{Verbatim}[xleftmargin=\codeindent,fontsize=\small]
void matrix_multiply(
  float* x, int x_rows, int x_cols, float* y, int y_cols, float* out) {
  for(int i = 0; i < x_rows; i++) {
    for(int j = 0; j < y_cols; j++) {
      for(int k = 0; k < x_cols; k++) {
        out[i][j] += x[i][k] * y[k][j]; }}}}
\end{Verbatim}

Indeed, the inner loop body is almost identical with what we wrote in Dex, and has the same benefit of explicitness about how the dimensions of the input arrays are handled.
However, by statically inferring the loop bounds, Dex is able to eliminate almost all of the syntactic noise---even more than the foreach syntax available in Java and C++.
On top of that, Dex's effect system exposes the parallelism available in this code, making automatic parallelization relatively straightforward (see \Cref{sec:parallelism-traversal}), and Dex's AD system breaks neither the parallelism nor the total work when differentiating through such computations (see \Cref{sec:ad-lessons}).
% \DV{Perhaps here we also need to say that the Dex compiler is not an optimizing compiler; it rather
% allows programmers to very easily write optimized code without requiring heroic efforts?}

\section{The Dex Language}
\label{sec:formal-language}

We now transition to describing Dex more formally and in more detail.
The structure of the rest of the paper mirrors the architecture of the Dex compiler:
\begin{enumerate}
\item Dex parses the surface language, elaborates syntactic sugar, and infers omitted type annotations, producing core IR.
We describe the core IR in \Cref{sec:core-ir}; we do not discuss parsing and desugaring since they are standard.
\item We give the type system in \Cref{sec:formal-type-system}.
Type inference uses a bidirectional algorithm insipred by \citet{peytonjones2007practical}, but we omit discussing it in the interest of space.
\item Dex simplifies (\Cref{sec:defunctionalization}) the core IR into a restricted subset that omits higher-order functions.
\item As part of simplification, Dex differentiates (\Cref{sec:ad}) all functions that appear as arguments to differentiation operators---the compiler treats AD as a subroutine of simplification.
This is both logical, since differentiation is a higher-order function itself, and convenient, since it permits Dex's AD to consume post-simplification IR but emit core IR, and rely on continuing simplification to clean up after it (e.g., see \Cref{sec:capturing-scoped-intermediates}).
\item Dex optimizes the simplified and differentiated IR with standard techniques and generates LLVM bytecode for final compilation for CPU or GPU (\Cref{sec:performance}).
Parallelism is extracted automatically (\Cref{sec:auto-parallel}).
\end{enumerate}

\begin{figure}[t]\scriptsize
\begin{tabular}{cc}
\begin{tabular}{r@{~~}r@{~~}ll}
  \multicolumn{3}{l}{Values (including types)} \\
  $v, \tau$ & $::=$ & $x$ & variable \\
  & $\mid$ & $l$               & literal \\
  & $\mid$ & \ttt{Type} $\mid$ \ttt{Unit} $\mid$ \ttt{Int} $\mid$ \ttt{Float} & base types \\
  & $\mid$ & $\fin v$   & finite index set \\
  & $\mid$ & $\tau \rightarrow {\epsilon}~\tau $   & function type \\% (with effect $\epsilon$)
  & $\mid$ & $\tau \Rightarrow \tau $   & array type \\
  & $\mid$ & $\tau \times \tau$         & pair type \\
  & $\mid$ & $\ttt{Either}~\tau~\tau$   & sum type \\
  & $\mid$ & $\ttt{Ref}~\tau~\tau$   & reference type \\
  & $\mid$ & $\lam x \tau e$   & function \\
  & $\mid$ & $\view x \tau e$  & table view \\
  & $\mid$ & $(v,v)$  & pair constructor \\
  & $\mid$ & $\ttt{Left}~\tau~v \mid \ttt{Right}~\tau~v$  & sum type constructors \\
%   & $\mid$ & $\case v {v~x} {v~x}$         & value case expression \\
  & $\mid$ & $\ttt{case}~v~\ttt{of}~
                         \ttt{Left\phantom{-} $x$} \rightarrow v~x$         & value case expression \\
  & & $\phantom{\ttt{case}~v~\ttt{of}~}
                         \ttt{Right $x$} \rightarrow v~x$         &\\
  \multicolumn{3}{l}{Contexts} \\
  $E$ & $::=$
           & $\bullet$  & hole\\
  & $\mid$ & $\letx {x:\tau} {e} E$  & let context \\
\end{tabular}
&
\begin{tabular}{r@{~}r@{~}ll}
  \multicolumn{3}{l}{Expressions} \\
  $e$ & $::=$
           &  $v$                  & value \\
  & $\mid$ & $\letx {x:\tau} {e_1} {e_2}$  & let expression \\
  & $\mid$ & $v~v$                 & function application \\
  & $\mid$ & $v.v$                 & table indexing \\
  & $\mid$ & $\forexpr x \tau e$   & table builder \\
  & $\mid$ & $\ttt{fst}~v \mid \ttt{snd}~v$       & pair projections \\
%   & $\mid$ & $\case v e e$         & case expression \\
  & $\mid$ & $\ttt{case}~v~\ttt{of}~
                         \ttt{Left\phantom{-} $x$} \rightarrow e$         & case expression \\
  & & $\phantom{\ttt{case}~v~\ttt{of}~}
                         \ttt{Right $x$} \rightarrow e$         &\\
  & $\mid$ & $v~!~v$               & reference slicing \\
  & $\mid$ & $\ttt{runState}~v~v \mid \ttt{get}~v \mid \ttt{put}~v~v$   & \ttt{State} operations \\
  & $\mid$ & $\ttt{runAccum}~v \mid v~\ttt{+=}~v$   & \ttt{Accum} operations \\
  & $\mid$ & $v~\ttt{+}~v \mid v~\ttt{*}~v$ & arithmetic operations\\
  & $\mid$ & \ttt{linearize}~v~v & linearization \\
  & $\mid$ & \ttt{transpose}~v~v & transposition \\
  \\
  \multicolumn{3}{l}{Effects} \\
  $\epsilon$ & $::=$
             &  $\ttt{Pure}$                & no effects \\
  & $\mid$   &  $\ttt{State}~\tau, \epsilon$   & state       effect \\
  & $\mid$   &  $\ttt{Accum}~\tau, \epsilon$   & accumulator effect \\
\end{tabular}
\end{tabular}
\caption{Dex core IR.  Dex's type system is value-dependent; here we distinguish $\tau$ and $v$ only to hint whether a given value is expected to have type \texttt{Type} or any type, respectively, as they are not actually different in the Dex grammar.
%Variable $x$ is scoped to $e_2$ in let expressions and to each branch of case expressions.
}
\label{fig:core-language-latex}
\end{figure}

\subsection{Core Dex IR}
\label{sec:core-ir}

While the full Dex language supports many features (e.g. algebraic data types, dependent pairs, type classes, implicit arguments, automatically synthesized arguments, etc), we focus our exposition here on the essential core of the language.
\Cref{fig:core-language-latex} presents a simplified version of the core intermediate representation (IR).
This is the central data structure in the Dex compiler.

The core IR is very close to the surface language, although it is somewhat compressed and requires explicit type annotations on all binders.
Incompletely annotated Dex surface terms are elaborated into this representation with a bidirectional type inference algorithm inspired by \citet{peytonjones2007practical}.

Most of the core language has fairly standard semantics. We have already discussed the \ttt{for} expression (\Cref{sec:typed-indexing}) and the effect system (\Cref{sec:effects}), and we will introduce the \ttt{view} expression in \Cref{sec:function-array-duality} (which provides an alternate way to construct values of array types).
%Most of the core language has fairly standard semantics.
% Since have already discussed the \ttt{for} expression (\Cref{sec:typed-indexing}) and the effect system (\Cref{sec:effects}),
%it just remains to describe the \emph{table view} abstraction.  The intent is to model array views or slices that do not need to copy their source buffer.
% A table view expression
% $$ \ttt{view i:n.expr} $$
% is similar to \verb|for i:n.expr| in that it constructs an object of the same type \verb|n=>a| and containing the same elements (produced by \ttt{expr}).
% However, it is like the lambda abstraction \verb|\i:n.expr| in that its body is evaluated lazily, each time the result is indexed, rather than eagerly as with \ttt{for}.
% Note, however, that because array indexing is expected to not carry any effects, the body of a \texttt{view} has to be pure.
% As we will see later (\Cref{sec:defunctionalization}, rule \tsc{SFor}), having such a lazy array type is useful for creating ``array coercions'' that will be used throughout the simplification pass.
%
One notable limitation of the IR is that \texttt{let} bindings in Dex are non-recursive, and at the moment there is no way to express recursion in the language.
While this limitation might seem significant, in our experience with Dex, the \texttt{for} iterator abstraction provides enough expressive power for quite a wide range of applications.
Also, compared to general recursion, \texttt{for} is both quite familiar to the scientific computing community and much easier to compile to the flat parallelism required by modern hardware accelerators.

Contexts, denoted $E$, represent sequences of let bindings with a hole in the place of the final result. These are used during the simplification and linearization passes, and are not a part of the surface language.
We write $E[e]$ for completing a context $E$ with an open term $e$ to obtain an expression.
Contexts may also be composed, which we write $E_1 \circ E_2$, by inserting $E_2$ into the hole of $E_1$; this concatenates the let bindings.

\begin{figure}\footnotesize
\renewcommand{\arraystretch}{3}
\begin{tabular}{cc}
$\Infer{TypeFor}
  {     \epsilon,~x\ann\tau_1,~\Gamma\vdash e : \tau_2
  \quad \vdash_{\ttt{IdxSet}} \tau_1}
  {\epsilon,~\Gamma\vdash (\forexpr x {\tau_1} e): (\tau_1\Rightarrow\tau_2)}$
& $\Infer{TypeView}
  {     \ttt{Pure},~x\ann\tau_1,~\Gamma \vdash e :  \tau_2
  \quad \vdash_{\ttt{IdxSet}} \tau_1}
  {\epsilon,~\Gamma \vdash (\view x {\tau_1} e) : (\tau_1 \Rightarrow \tau_2)}$
\\$\Infer{TypeIndex}
  {\Gamma\vdash v_1:(\tau_1 \Rightarrow \tau_2) \quad
   \Gamma\vdash v_2:\tau_1}
  {\epsilon,~\Gamma\vdash (v_1.v_2):\tau_2}$
& $\Infer{TypeSlice}
  {\Gamma\vdash v_1 : \ttt{Ref}~h~(\tau_1\Rightarrow\tau_2) \quad
   \Gamma\vdash v_2 : \tau_1}
  {\epsilon,~\Gamma\vdash (v_1 ! v_2 ) : \ttt{Ref}~h~\tau_2}$
\\$\Infer{TypeGet}
  {\Gamma \vdash v : \ttt{Ref}~h~\tau }
  {\ttt{State}~h,~\epsilon,~\Gamma\vdash \ttt{get}~v : \tau}$
& $\Infer{TypePut}
  {\Gamma \vdash v_1 : \ttt{Ref}~h~\tau \quad
   \Gamma \vdash v_2 : \tau }
  {\ttt{State}~h,~\epsilon,~\Gamma\vdash \ttt{put}~v_1~v_2 : \ttt{Unit}}$
\\
\multicolumn{2}{c}{
$\Infer{TypeRunState}
  {\Gamma \vdash v_1 : \tau_1 \quad
   \Gamma \vdash v_2 : (\ttt{Ref}~h~\tau_1 \rightarrow (\ttt{State}~h,~\epsilon)~\tau_2) \quad
   \vdash_{\ttt{Data}} \tau_1}
  {\epsilon,~\Gamma\vdash \ttt{runState}~v_1~v_2 : (\tau_2 \times \tau_1) }$
}
\end{tabular}
\caption{Subset of Dex's typing rules for arrays and effects;
the remaining rules can be found in
\ifextended \Cref{fig:typerules-full}
and \Cref{fig:typerules-constraints} in \Cref{sec:full-rules}.
\else the supplementary material.
\fi For type checking, we model effects as capabilities in the spirit of \citet{brachthauser2020effects}.}
\label{fig:typerules}
\end{figure}

\subsection{Type system}
\label{sec:formal-type-system}
In Dex, we use a form of value dependent types \cite{swamy2011valuedependent}, and hence our core language is separated into two syntactic categories: values and expressions.
Values approximately correspond to the fully reduced terms that can be implicitly lifted to appear in types.%
\footnote{One notable exception is the ``value case'' construct, which is generally not considered work-free.  We include it as a value on a technicality: we need it to represent the result of simplifying a \ttt{case} of function type (see rule \tsc{SCase} in \ifextended \Cref{fig:defunctionalization-full}\else the extended version\fi).  We do not recognize ``value case'' values when parsing user code, defaulting to \ttt{case} expressions.  In particular, value case does not occur in types.} %
So, while we do not allow arbitrary expressions in types, we do allow arbitrary \emph{expression results}, provided they are bound to a variable.
This makes type checking quite straightforward, as type equality remains syntactic (up to alpha-equivalence).
The downside is a small loss in precision: when two terms that reduce to the same value are bound to different variables, this can cause type mismatches between subsequent values that lift those variables into their types.
%The downside is that two terms that reduce to the same value, but are bound to different variables, can cause type mismatches, so there is a small loss in precision.
In typical Dex programs, however, this implicit lifting is used only for array shapes, and those are fortunately usually each lifted just once in the program, diminishing the importance of this drawback.
% \DV{This calls for an example somewhere.}

The most interesting typing rules for the core language are presented in \Cref{fig:typerules}.
The typing judgement $\epsilon, \Gamma \vdash e: \tau$ can be read as ``given the capability to perform effects $\epsilon$, expression $e$ evaluated in an environment with variables of types $\Gamma$ produces a result of type $\tau$.''
Note that effects can also appear in types, but only on the right-hand side of a function type $\tau_1 \rightarrow \epsilon\ \tau_2$.

% Note that the type of \texttt{Type} is also \texttt{Type}.
% Dex doesn't implement a hierarchy of universes.
% \apc{TODO: Explain why we can get away without it.}

\subsection{Duality of functions and arrays}
\label{sec:function-array-duality}

We chose \texttt{=>} as the array type constructor for its similarity to the function type constructor \texttt{->}.
Indeed, functions and arrays are almost perfectly analogous. Both are language constructs with abstraction and application, and both have the same reduction rule:
 \verb|(\x. expr) y| reduces to \verb|[x->y]expr| just as well as \verb|(for x. expr).y| does (at least when \ttt{expr} is pure).
The only real difference is in \emph{when} the abstraction body is evaluated.
In the case of functions, this happens at each application site, whereas in the case of \ttt{for} expressions it happens eagerly when the array is first defined.
% TODO: Consider if we have space!
% \DV{Interesting: but it needn't! One could keep the "lazy" form of arrays. Perhaps
% worth noting that this is what F-smooth would do. But then you can say: well if you want a lazier form
% of arrays, just use functions, because Dex is {\bf higher order and functions are fist class}! Nicely
% motivating the higher-orderness. And we can even give an example, I would strongly suggest.}

Arrays are in fact just a representation for a fully memoized function: the application of an array computes the element by looking it up in memory.
Conversely, functions can be seen as ``lazy'' or ``compressed'' arrays that compute the elements just-in-time as they are requested.
Functions also do not require their argument type to be enumerable, whereas precomputing an array forces that requirement on us.
Likewise, if the body of a function has an effect, that effect occurs when the function is called, whereas a \ttt{for} expression executes all the effects for all the iterations immediately (when relevant, the order defined to be is the element order of the index set).

In some cases, it is useful to produce a value of type \verb|n=>a| without fully memoizing it; this can be used to model \emph{views} or \emph{slices} of an existing (memoized) array without unnecessary memory copies. We thus include in the core IR a table view expression
$$ \ttt{view i:n.expr}, $$
which has the same type as \verb|for i:n.expr| but is evaluated lazily, each time the result is indexed, rather than eagerly.
Note, however, that because array indexing is expected to not carry any effects, the body of a \texttt{view} has to be pure.
As we will see later (\Cref{sec:defunctionalization}, rule \tsc{SFor}), having such a lazy array type is useful for creating ``array coercions'' that will be used throughout the simplification pass.
%it just remains to describe the \emph{table view} abstraction.  The intent is to model array views or slices that do not need to copy their source buffer.
% A table view expression
% $$ \ttt{view i:n.expr} $$
% is similar to \verb|for i:n.expr| in that it constructs an object of the same type \verb|n=>a| and containing the same elements (produced by \ttt{expr}).
% However, it is like the lambda abstraction \verb|\i:n.expr| in that its body is evaluated lazily, each time the result is indexed, rather than eagerly as with \ttt{for}.
% Note, however, that because array indexing is expected to not carry any effects, the body of a \texttt{view} has to be pure.
% As we will see later (\Cref{sec:defunctionalization}, rule \tsc{SFor}), having such a lazy array type is useful for creating ``array coercions'' that will be used throughout the simplification pass.

One might imagine taking this analogy to the logical extreme and dispensing with the distinction between functions and arrays entirely, at least for pure expressions when the evaluation order matters less.
We have not seen a system take that approach successfully, and in Dex, we instead leave the choice of representation in the hands of the programmer.

\subsection{Types of array indices}
\label{sec:index-sets}

As we noted when we introduced the \texttt{for} expression, not all types can be used as indices.
We call every type that can be so used an \emph{index set}.
In this section we summarize the requirements for an index set, and briefly discuss how we discharge them in Dex.

Firstly, each index set is required to have a finite number of members, as otherwise we couldn't represent the array in finite memory.
Secondly, because index sets also define iteration order in \texttt{for} expressions, which can have side effects, we need to be able to enumerate the members of each such type in a fixed total order.
Hence, we also require each index set to specify a bijection to integers between 0 and the size of the index set (exclusive).
Both of those requirements can be conveniently summarized in a Haskell-style type class, here presented using Dex's syntax for type classes (which we do not discuss further, due to space concerns):
\begin{Verbatim}[xleftmargin=\codeindent,fontsize=\small]
interface IndexSet a where
  size        : Int
  ordinal     : a -> Int   -- returns integers between 0 and size - 1
  fromOrdinal : Int -> a   -- partial and only defined on valid ordinals
\end{Verbatim}

Even though technically the type of, e.g., 64-bit integers or IEEE floating-point values satisfies those requirements, we do not consider them to be index sets, because it is unlikely that a user would want to have an array with as many as $2^{64}$ entries.
Instead, the basic index set type provided by Dex is \verb|Fin : Int -> Type|.
It is a builtin type constructor which guarantees that \texttt{Fin n} has exactly \texttt{n} members.
It is often convenient to think about it as a prefix of the natural numbers up to \texttt{n}, although there are no literals of this type available by default.
Valid instances can only be obtained from \texttt{for} binders or \texttt{fromOrdinal}.

For example, a $5 \times 4$ matrix of floats in Dex can be typed as
\verb|(Fin 5)=>(Fin 4)=>Float|.
But \texttt{Fin} is a regular function that can accept arbitrary integer values, not just literals.
In particular, because the Dex type system implements a form of value-dependent types \cite{swamy2011valuedependent}, it is possible to represent arrays of statically unknown size:
\begin{Verbatim}[xleftmargin=\codeindent,fontsize=\small]
n = sum $ for i:(Fin 100). ordinal i * ordinal i
x : (Fin n)=>Float = for i. 1.0
\end{Verbatim}

Dex also allows the definition of richer index sets than \texttt{Fin}, which we found useful for expressing a wide range of numerical algorithms.
Space prevents us from covering the possibilities thoroughly here, but we highlight tuples as one particularly good example.
A tuple of index sets is a valid index set if and only if all its components are valid index sets.
Tuples let us capture a type-safe ``flattening'' and ``unflattening'' transformation as a form of currying for arrays:
\begin{Verbatim}[xleftmargin=\codeindent,fontsize=\small]
x  : n=>m=>Float = ...
y  : (n & m)=>Float = for (i,j). x.i.j
x' : n=>m=>Float = for i j. y.(i,j)
\end{Verbatim}
Now, any index-polymorphic Dex library function of type, say, \verb|g: (a=>Float) -> (a=>Float)| is usable with data of type \texttt{x} as
\begin{Verbatim}[xleftmargin=\codeindent,fontsize=\small]
g  : (a=>Float) -> (a=>Float)
gx : n=>m=>Float =
  y = g (for (i,j). x.i.j)
  for i j. y.(i,j)
\end{Verbatim}
This captures one of the most common uses for the \texttt{reshape} operation common to bulk array programming, while both preserving static information about array sizes, and not requiring the type system to solve systems of Diophantine equations to check which \texttt{reshape}s are valid.

% \subsubsection{Dependent arrays}
% \label{sec:dependent-arrays}

% Identifying functions and arrays in a dependently typed language, leads to a natural research question: if that functions allow dependencies between the types of their arguments, what would happen if we allowed dependencies between the types of array indices as well?
% While we have not completed this investigation in full, we did want to mention here that it does leads to quite promising results.

% Assume that \texttt{i} is a member (i.e. a value) of a valid index set \texttt{n}.
% Now, consider a type written as \texttt{..i}.
% We assume that this type is a valid index set as well, and it contains all indices that belong to \texttt{n}, but only up to \texttt{i} (inclusive).

% Now, consider the array type \texttt{i:n=>(..i)=>a}.
% Hence, introducing dependencies between array dimensions lets us express \emph{non-rectangular arrays}.
% So far the only work we are aware of that has explored this direction is \cite{pizzuti2019positiondependent}, although our hope is to bring more structure to those dependencies through a number of carefully designed index sets.

\section{Simplification to first-order programs}
\label{sec:defunctionalization}
\newcommand\eminor{E^\flat}
\newcommand\vminor{v^\flat}
\newcommand\tminor{\tau^\flat}
\newcommand\freevars[1]{{\tt free}(#1)}
\newcommand\boundvars[1]{{\tt bound}(#1)}
\newcommand\binders[1]{{\tt binders}(#1)}
%\DV{This is {\bf not} defunctionalization. This is just aggressive inlining/partial evaluation,
%relying on the sub-formula property and with extra care to avoid duplication of work. You can see
%the same ideas in \url{https://homepages.inf.ed.ac.uk/slindley/papers/dpia-draft-july2018.pdf},
%Section 4.1. There however they translate directly to imperative code whereas here we simply
%aggressively eliminate lambdas, so we are more reminiscent of their previous work:
%\url{https://research.chalmers.se/en/publication/234804}. The essence is that a higher-order
%program with only first-order primitives in its context and a first-order return type can be, %provably,
%through partial evaluation converted into an entirely first-order program. But let us not call this
%here defunctionalization. (Defunctionalization is typically treating function types as forms of
%objects, and replaces application with a single primitive {\tt apply} function (the only function
%remaining in the program), see \url{https://en.wikipedia.org/wiki/Defunctionalization}}

Higher-order functions are a big part of what makes functional programming so
much fun, but it's tricky to compile them to efficient machine code.
Accelerators do not always support function calls, or oftentimes have a high overhead penalty associated with them. In addition, AD becomes especially tricky with higher-order functions \citep{manzyuk2019perturbationconfusion, ritchie2021higherorderad}.
To avoid these problems we apply a full {\em simplification pass} in the Dex compiler prior to AD and code generation. After this pass, the only functions left are monomorphic first-order top-level functions that are immediately and fully applied at their use sites in a program, provided that this program returns a non-function type. Similar normalization procedures are not uncommon in array languages~\cite{hovgaard2018high, najd+:everything-old}, and are inspired by cut elimination in formal logic.

Figure~\ref{fig:defunctionalization} presents the details of this pass, with novel aspects tailored to Dex's constructs (arrays and effect handlers) and performance considerations.
A point of departure from previous work is that the pass allows returned values to be of function type, which occurs when a first-order function uses higher-order constructs internally.
We are also able to get much more mileage out of this kind of simplification because Dex has no constructs for defining recursive functions.

The judgement form
\[  e \defunc E^d, v \]
represents the conversion of an expression $e$ in Dex core IR to a {\em simplification context} $E^d$
that contains all the computation that needs to be performed before $e$ reaches a value form.
Here $v$ is an (ordinary) value in Dex core IR. Figure~\ref{fig:defunctionalized-language} presents
the subset of Dex's core IR that can be reached through these simplification contexts.
% Contexts are simply sequences of let-bindings that bind simplified expressions $e^d$. These
The expressions $e^d$ that may appear in a simplification context
are very much like the expressions of Core but only contain values that come from a syntactic
subset of Dex values, which we denote with $v^d$. The post-simplification IR in particular
does not include functions or function types.

\begin{figure}[t]\scriptsize
\begin{tabular}{cc}
\begin{tabular}{r@{~~}r@{~~}ll}
  \multicolumn{3}{l}{Values} \\
  $v^d$ & $::=$ & $x$ & variable \\
  & $\mid$ & $l$               & literal \\
  & $\mid$ & $\view x {\tau^d} b^d$  & table view \\
  & $\mid$ & $(v^d,v^d)$  & pair constructor \\
  & $\mid$ & $\ttt{Left}~\tau^d~v^d \mid \ttt{Right}~\tau^d~v^d$
              & \ttt{Either} constructors \\
  \multicolumn{3}{l}{Types} \\
  $\tau^d$ & $::=$ &
  \ttt{Type} $\mid$ \ttt{Unit} $\mid$ \ttt{Int} $\mid$ \ttt{Float} & base types \\
  & $\mid$ & $\fin v^d$   & finite index set \\
  & $\mid$ & $\tau^d \Rightarrow \tau^d $   & table type \\
  & $\mid$ & $\tau^d \times \tau^d$         & pair type \\
  & $\mid$ & $\ttt{Either}~\tau^d~\tau^d$   & sum type \\
  & $\mid$ & $\ttt{Ref}~x~\tau^d$   & reference type \\
  \multicolumn{3}{l}{Blocks} \\
  $b^d$ & $::=$
           & $v^d$ & value \\
  & $\mid$ & $\letx {x\ann\tau^d} {e^d} {b^d}$ & let expression \\
\end{tabular}%
&%
\begin{tabular}{r@{~~}r@{~~}ll}
  \multicolumn{3}{l}{Expressions} \\
  $e^d$ & $::=$
           & $x.v^d$                 & table indexing \\
  & $\mid$ & $\forexpr x {\tau^d} b^d$   & table builder \\
  & $\mid$ & $\ttt{fst}~x \mid \ttt{snd}~x$       & pair projections \\
%   & $\mid$ & $\case x {b^d} {b^d}$ & case expression \\
  & $\mid$ & $\ttt{case}~x~\ttt{of}~
                         \ttt{Left\phantom{-} $x$} \rightarrow b^d$         & case expression \\
  & & $\phantom{\ttt{case}~v~\ttt{of}~}
                         \ttt{Right $x$} \rightarrow b^d$         &\\
  & $\mid$ & $x~!~v^d$               & reference slicing \\
  & $\mid$ & $\ttt{runState}~v^d~(\action h x {\tau^d} b^d)$   & \ttt{State} handler \\
  & $\mid$ & $\ttt{get}~v^d \mid \ttt{put}~v^d~v^d$   & \ttt{State} operations \\
  & $\mid$ & $\ttt{runAccum}~(\action h x {\tau^d} b^d)$   & \ttt{Accum} handler \\
  & $\mid$ & $v^d~\ttt{+=}~v^d$   & \ttt{Accum} update \\
  & $\mid$ & $v^d~\ttt{+}~v^d \mid v^d~\ttt{*}~v^d$ & arithmetic operations \\
  \\
  \multicolumn{3}{l}{Contexts} \\
  $E^d$ & $::=$
           & $\bullet$  & hole\\
  & $\mid$ & $\letx {x:\tau^d} {e^d} E^d$  & let context \\
\end{tabular}
\end{tabular}
\caption{Post-simplification Dex IR, a first-order subset of the core IR given in \Cref{fig:core-language-latex}.}
\label{fig:defunctionalized-language}
\end{figure}

\begin{figure}\footnotesize
\renewcommand{\arraystretch}{3}
\begin{tabular}{c}
\framebox{$e \defunc E^d , v$} \\
 \Infer{SVal}{~~}{v \defunc \bullet, v} \qquad
  \Infer{SExpr}
  {e^d : \tau^d\qquad x\ \mathrm{fresh}}
  {e^d \defunc \letx{x:\tau^d}{e^d}\bullet , x } \quad\quad
  \Infer{SApp}
  {\subst{x}{v}{e} \defunc E^d, v'}
  {(\lam x \tau e)~v \defunc E^d, v'} \quad\quad \\
\Infer{SLet}
  {             e_1  \defunc E^d_1, v_1  \qquad
   \subst{x}{v_1}{e_2} \defunc E^d_2, v_2}
  {\letx {x:\tau} {e_1} {e_2} \defunc E^d_1\circ E^d_2, v_2} \quad\quad
\Infer{SView}
  {\subst{x}{v}{e} \defunc E^d, v'}
  {(\view x \tau e).v \defunc E^d, v'} \\
\Infer{SFor}
  {e \defunc E^d, v  \qquad  \binders{E^d}\vdash v \triangleright \overline{x}^{1..n}
  \qquad x \notin \freevars{x_1,\ldots,x_n}
  \qquad y~\mathrm{fresh}}
  {\forexpr x \tau e \defunc
        (\letx {y} {\forexpr x \tau E^d[(x_1,\ldots,x_n)]} \bullet) ,
        ( \view x \tau \letx {(x_1,\ldots,x_n)} {y.x} v )  }\\
\Infer{SLinearize}
     {e\defunc E_1^d,~v_1  \quad
      \linreify{\Gamma}{x}{E_1^d[v_1]} e' \quad
      e' \defunc E_2^d,~v_2
         }
 {\ttt{linearize}~(\lam x \tau e)~v\defunc \subst{x}{v}(E_2^d,~v_2)}
 \\
\Infer{STranspose}
     {e\defunc E_1^d,~v_1  \quad
      \transrule{x\rightarrow r}{E_1^d[v_1]} {v_t} e' \quad
      \ttt{yieldAccum}~(\action {h} r {\ttt{Ref}~h~\tau} e') \defunc E_2^d,~v_2
      \quad r,h~~\mathrm{fresh}
         }
 {\ttt{transpose}~(\lam x \tau e)~{v_t}\defunc E_2^d,~v_2}

 \vspace{-1.5em}
 \\
\framebox{$\overline{x{:}\tau} \vdash v \triangleright \overline{y}$} \\
\Infer{Empty}{~~}{\cdot \vdash v \triangleright \emptyset} \qquad
\Infer{Used}
    { \overline{x{:}\tau_x} \vdash v \triangleright \overline{y} \qquad
        x_1 \in \freevars{v} \\\\\ \freevars{\tau_1} \cap \overline{y} = \emptyset}
    {(x_1{:}\tau_1),\overline{x{:}\tau} \vdash v \triangleright x_1,\overline{y}} \qquad
\Infer{NotUsed}
    { \overline{x{:}\tau_x} \vdash v \triangleright \overline{y} \qquad
        x_1 \notin \freevars{v}}
    {(x_1{:}\tau_1),\overline{x{:}\tau} \vdash v \triangleright \overline{y}}
\end{tabular}
\vspace{-0.5cm}
\caption{Subset of simplification rules. See
\ifextended \Cref{fig:defunctionalization-full}
\else the extended version
\fi
for the remaining rules. $\mathcal{L}$ and $\mathcal{T}$ are defined in \Cref{fig:linearize} and \Cref{fig:transpose} respectively.
}
\label{fig:defunctionalization}
\vspace{-0.5cm}
\end{figure}

Simplification is semantics-preserving and non-work-increasing, in the sense that if $e \defunc E^d, v$ then completing $E^d[v]$ produces a term operationally equivalent to $e$ which does not introduce any more work.
% \AXCH{Operational equivalence is a term of art.  Somebody check the preceding sentence.}
% I checked, it's fine.
We inline let bindings (rule \tsc{SLet}) and beta reduce wherever possible
(rules \tsc{SApp} and \tsc{SView}). To avoid duplicating runtime work, we only
substitute let- and lambda- bound variables with values $v$. We emit let bindings for expressions we want to evaluate at run-time (rule \textsc{SExpr}), adding them
to the context, $E^d$.

The difficult part is simplification through control constructs like \texttt{for}. What
happens if our source program builds a table of functions, like the following?
\begin{Verbatim}[xleftmargin=\codeindent,fontsize=\small]
for i.
  y1 = f1 xs.i  -- f1 is an expensive function
  y2 = f2 y1    -- f2 is an expensive function
  \z. y1 + y2 + z
\end{Verbatim}
Each function in the table is different, but they all share the same code: \texttt{\textbackslash z.
y1 + y2 + z}. The only meaningful difference is the run-time value of the variables
\texttt{y1} and \texttt{y2}, accessed from the function's lexical scope. These might be expensive to compute, so we don't want to just inline the table definition at its indexing sites.
Instead, the simplified context captures the variables \texttt{y1} and
\texttt{y2} (or rather their simplified counterparts) as a tuple, turning the table of
functions into a table of data. The value we produce is a \ttt{view}
that indexes into the table of data to reconstitute each function:

\mbox{}\vspace{-0.15cm}\\
\begin{minipage}{0.35\textwidth}
\begin{Verbatim}[xleftmargin=\codeindent,fontsize=\small]
-- Simplified context
let xs = for i.
  y1 = f1 xs.i
  y2 = f2 y1
  (y1, y2)
in
\end{Verbatim}
\end{minipage}%
\begin{minipage}{0.5\textwidth}
\begin{Verbatim}[xleftmargin=\codeindent,fontsize=\small]
-- Residual value reconstructing original type
view i.
  (y1, y1) = xs.i
  \z. y1 + y2 + z
\end{Verbatim}
\end{minipage}

\mbox{}\vspace{-0.25cm}\\

This transformation is handled by the rule \textsc{SFor}. %, which we reproduce below:
% Rule \textsc{SFor} achieves this. We replicate and explain the rule below:
% {\footnotesize%
% \begin{align*}
% & \Infer{SFor}
%   {e \defunc E^d, v  \qquad  \binders{E^d}\vdash v \triangleright \overline{x}^{1..n}
%   \qquad x \notin \freevars{x_1,\ldots,x_n}
%   \qquad y~\mathrm{fresh}}
%   {\forexpr x \tau e \defunc
%         (\letx {y} {\forexpr x \tau E^d[(x_1,\ldots,x_n)]} \bullet) ,
%         ( \view x \tau \letx {(x_1,\ldots,x_n)} {y.x} v )  }
% \end{align*}%
% }%
First we recursively simplify the body of the expression $e$ into a context $E^d$ and a residual
value $v$. Next, the judgement $\binders{E^d}\vdash v \triangleright \overline{x}^{1..n}$ calculates
to a first approximation calculates the free variables of $v$ that are bound by $E^d$. The context
we return, $(\letx {y} {\forexpr x \tau E^d[(x_1,\ldots,x_n)]} \bullet)$, effectively binds a
table of tuples for the context-bound variables. Finally the value we return can
deconstruct, for every element in that table, a tuple of these values and use them in $v$.

Returning to the $\overline{x{:}\tau} \vdash v \triangleright \overline{y}$ judgement, we observe that
the judgement is conservative to ensure that the variables collected ($\overline{y}$) have types that
do not depend on other variables and hence can be tupled together in a {\em non-dependent} pair. With
more dependency the situation is more complex. Consider the simplification of:
\begin{verbatim}
  for x.
     n = ...
     xs = for i:(Fin n). ...
     \w. sum xs
\end{verbatim}
Although {\tt n} does not appear in the closure as a free variable, its type mentions the variable
that does appear. As it stands Dex will not perform simplification of this program, and report an error.
The solution to this is to introduce {\em dependent pairs} when floating definitions out of bodies
(rule \tsc{SFor}, and rules \tsc{SCase, SRunAccum, SRunState} in
\ifextended
\Cref{fig:defunctionalization-full}
\else
the extended version
\fi) and convert our list of free variables into {\em telescopes} tracking their dependencies. We leave a full formal treatment of this idea and the implementation as future work.

%Another delicate point is the fact that rules \tsc{SRunState} and \tsc{SRunAccum} consider
%the bound variable $x$ of {\em reference type} when calculating the variables that must be returned
%from the state handler. In particular we see preconditions of the form:
%$(x{:}\tau),\binders{E^d} \vdash v' \triangleright \overline{x}^{1..n}$. However, if the value
%$v'$ were to actually read or modify the passed-in reference then the program could not have been well-typed to start with, because the state thread would leak in the returned type. Perhaps though the program does not really read or modify the reference but simply throws it away and returns some other value. The rule permits this binding only to cover this last set of corner-case programs. We
%conjecture that our simplification rules can ensure these programs could be simplifiable to not
%mention the reference variable at all and that we could drop $(x{:}\tau)$ from the variable calculation
%precondition of these rules.

Finally, this pass also serves as a
monomorphization pass. Polymorphic functions are just ordinary functions that
take a type as an argument. We inline and then beta-reduce these functions at
their use-sites, just as we do for higher-order functions.

\section{Automatic differentiation}
\label{sec:ad}

Efficient automatic differentiation (AD) is a keystone for machine learning, and is increasingly important for broader numerical computing.
% Argument that defunctionalization is total is subtle, and relies on type-correctness and type class constraints.  Wait for reviewers to ask for it.
For that reason, Dex is designed around efficient AD.
We now explain how Dex implements AD, and then consider several ways in which the goal of efficient and complete AD interacted with the design of Dex in \Cref{sec:ad-lessons}.
For a more detailed exposition of automatic differentiation we refer to a survey by \citet{baydin2017adsurvey}.

\subsection{Linearization}
\label{sec:linearization}

The semantics of automatic differentiation of programming language functions are defined in terms of differentiation of mathematical functions.
Mathematically, for a sufficiently nice function $f: \R^n \to \R^m$, its derivative $\partial f$ evaluated at a point $x \in \R^n$ is the linear map $\partial f(x): \R^n \to \R^m$, uniquely defined by
\[
    f(x + v) = f(x) + \partial f(x)(v) + o(\|v\|), \qquad \forall v \in \R^n.
\]
We can think of $v$ as representing a perturbation to the input $x$, and $\partial f(x)(v)$ as representing the corresponding change in the output, to first order.
When speaking of automatic differentiation, $x$ and $f$ are called the \emph{primal} input and computation,  respectively, and $v$ is called the \emph{tangent}; the computations on $v$ that occur inside $\partial f(x)$ are similarly called tangent computations.
It can be convenient to identify $\partial f(x)$ with an $\R^{m \times n}$ matrix of partial derivatives called the Jacobian,
but representing $\partial f(x)$ as a function lets us capture the sparsity that arises from the data flow graph of the implementation of $f$.

To model this mathematical definition computationally, Dex provides a built-in function
\[
\verb|linearize : [VectorSpace a, VectorSpace b] (a -> b) -> a -> (b & a -o b)|
\]
In words, given a function of type \texttt{a -> b} representing a mathematical function $f$, and an input of type \ttt{a} representing a point $x$ in $f$'s domain, \texttt{linearize} produces an output of type \texttt{b} representing $f(x)$, and a \emph{structurally linear} function of type \texttt{a -o b} representing $\partial f(x)$.%
\footnote{This type signature puns the type of the tangent space for \texttt{a} with \texttt{a} itself.  That's reasonable when \texttt{a} is a fixed-shape structure of real numbers like \texttt{n=>Float}.  We do permit $f$ to use things like integers internally, for which we have to define an implementation-internal type function $\tangentof \cdot$, below.}
We do not formalize the concept of structural linearity here,\footnote{And Dex currently does not enforce it---the linear arrow \texttt{-o} is provided purely as documentation and is treated equivalently to~\texttt{->}. Verifying structural linearity as a typing judgement would be an interesting future extension.}
but the intuition is that the implementation of $\partial f(x)$ should never compute any intermediates that are non-linear in the input to $\partial f(x)$.
As an example, a program such as \verb|g = \x:Float. (x * x) / x| is linear mathematically (as it can be simplified to the identity function assuming infinite precision) but not structurally, because it computes the non-linear term \texttt{x * x} as an intermediate value.

Beyond the mathematical specification, \texttt{linearize} in Dex is a well-behaved computational object:
\begin{itemize}
\item The computational cost of linearizing a function $f$ at a point $x$, and of applying the linearized function $\partial f(x)$, is bounded by a small constant multiple of the cost of applying $f$ to $x$.
%That is, evaluating \texttt{(snd (linearize f x)) v} should require a small multiple as many basic operations as evaluating \texttt{f x}.
\item The linearized function $\partial f(x)$ is structurally similar to $f$: it relies on the same effects, and therefore exposes the same degree of parallelism.
\item Every Dex function of suitable type can be linearized,%
\footnote{Notably, the \texttt{VectorSpace} constraint restricts $f$'s type to first-order.  AD of higher-order functions is subtle, \cite{manzyuk2019perturbationconfusion, ritchie2021higherorderad}, so Dex eschews it.  But note that $f$ is free to use higher-order functions internally---the restriction is only that $f$ cannot be higher-order itself.} %
including functions produced by or using linearization, allowing the computation of higher-order derivatives.
\end{itemize}
Linearization is an instance of \emph{forward-mode AD} \citep{griewank2008evaluating}, as in e.g.~\citet{elliott2018ad}.
%in a formulation first proposed by \citet{elliott2018ad}
%\MATT{I commented the above phrase out because I wasn't sure what it meant (i.e. what exactly was first proposed by Conal). Rephrased below.}
In Dex, we realize it in the form of a compile-time source transformation, with a subset of interesting rules outlined in \Cref{fig:linearize}.

There are two types of rules we consider.  First,
$$\linrule{\diffenv}{b^d}\linresult{E}{e_p}{e_t}$$
is the main elaboration used in the process of linearization.
Given a mapping $\diffenv$ from primal program variables to their respective types and tangents, it translates a (simplified) expression or block $b^d$ into three (core IR) elements: (1) a context $E$, (2) a primal expression $e_p$, and (3) a tangent expression $e_t$.
The invariant for $\mathcal{D}_\diffenv$ is that $E[e_p]$ is equivalent to $b^d$, and $E[e_t]$ computes the tangent corresponding to $b^d$ assuming the free variables of $b^d$ are given the tangent values in $\Delta$.

Second,
$$\linreify{\Gamma}{x_1, \ldots, x_n}{b^d}e$$
encodes an elaboration rule that linearizes a simplified block or expression $b^d$ with respect to some of its free variables $x_1, \ldots, x_n$, and reifies the result as a primal value and tangent function.
The result $e$ is of pair type.
The first component of $e$ is equivalent to $b^d$; and the second component of $e$ evaluates to a function that accepts tangent values for the $x_i$ and returns the corresponding tangent for $b^d$.
% For example, \texttt{linearize (\textbackslash x. e)} is simplified to a function of the primal $x$, whose body is constructed by $\linreifyop{\freevars{e}}{x}{e}$.

\Cref{fig:linearize} also includes a few supporting operations.
% The remaining operations in \Cref{fig:linearize} are support.
$\zeroat \tau$ constructs a zero value of the vector space instance associated with $\tau$.
Note that having the array shape as part of the type is crucial for this to be well-defined.
$\tangentof \tau$ maps the type $\tau$ to its \emph{tangent type} (intuitively the type of infinitesimal perturbations to $\tau$).
Tangent types necessarily have to be vector spaces, which is why we leave $\tangentof{\ttt{Either}~\tau_1~\tau_2}$ undefined (it is unclear which case should be used for zero).
This problem can be worked around by making the mapping to the tangent type depend on the \emph{value} instead of its type, in which case the type of the tangent would match the tangent type of the constructor used in the primal value.
We leave this extension for future work.

\begin{figure}\footnotesize
\begin{align*}
& \framebox{$\tangentof \tau$} \\
&       \tangentof{\ttt{Float}} = \ttt{Float}
  \qquad \tangentof{\ttt{Int}} = \ttt{Unit}
  \qquad \tangentof{\tau_1 \Rightarrow \tau_2} = (\tau_1 \Rightarrow \tangentof{\tau_2}) \\
&       \tangentof{(\tau_1~\times~\tau_2)} = (\tangentof{\tau_1}~\times~\tangentof{\tau_2})
  \qquad \tangentof{\ttt{Either}~\tau_1~\tau_2} = \text{Unsupported!} \\
& \framebox{$\Delta [v]$} \quad
   \Delta ::= x_1 \rightarrow v_1, \ldots, x_n \rightarrow v_n  ;~\Gamma \\
&        (\ldots, x\rightarrow v, \ldots;~\Gamma)[x] = v
  \qquad (\ldots;~x:\tau)[x] = \zeroat{\tangentof \tau}
  \qquad \Delta[l] = \zeroat{\tangentof \tau}~(\text{when}~l\ann\tau) \\
& \framebox{$\linreify{\Gamma_d}{x_1, \ldots, x_n}{b^d}e$} \\
&\Infer{LinReify}
   {\linrule{x_1\rightarrow t_1,\ldots,x_n\rightarrow t_n;\Gamma^d}{b^d}
       \linresult{E}{e_p}{e_t} \qquad
    t_1\ldots,t_n~\mathrm{fresh}}
   {\linreify{\Gamma_d}{x_1, \ldots, x_n}{b^d}
         E[(e_p,~\lamnoann {t_1\ldots t_n} e_t)] } \\
& \framebox{$\linrule{\diffenv}{b^d}\linresult{E}{e_p}{e_t}$} \\
& \Infer{LinBlockResult}
  {}{\linrule \diffenv v \linresult \bullet v {\diffenv[v]}} \\
& \Infer{LinLet}
  {\linrule{\diffenv}{e^d}\linresult{E_1}{e_{p_1}}{e_{t_1}} \qquad
   \linrule{x\rightarrow t,~\diffenv}{b^d}\linresult{E_2}{e_{p_2}}{e_{t_2}} \qquad
   t~\mathrm{fresh}
  }
  {\linrule{\diffenv}{\letx {x\ann\tau} {e^d} {b^d}}
       \linresult{(E_1 \circ (\letx {x\ann\tau} {e_{p_1}} {\bullet}) \circ E_2)}
                 {e_{p_2}}
                 {(\letx {t\ann\tangentof \tau} {e_{t_1}} {e_{t_2}})}} \\
& \framebox{$\linrule{\diffenv}{e_d}\linresult{E}{e_p}{e_t}$} \\
&\Infer{LinAdd}
  {}{\linrule{\diffenv}{v_1~\ttt{+}~v_2}
      \linresult{\bullet}{v_1~\ttt{+}~v_2}{\diffenv[v_1]~\ttt{+}~\diffenv[v_2]}} \\
&\Infer{LinMul}
  {}{\linrule{\diffenv}{v_1~\ttt{*}~v_2}
      \linresult{\bullet}{v_1~\ttt{*}~v_2}{
           ((         v_1~\ttt{*}~\diffenv[v_2]) ~\ttt{+}~
            (\diffenv[v_1]~\ttt{*}~        v_2)
            )}} \\
& \Infer{LinFor}
  {\linreify{i\ann\tau,~\Gamma}{x_1, \ldots, x_n}{b^d} e \qquad
   j~\mathrm{fresh}}
  {\linrule{x_1\rightarrow t_1, \ldots, x_n\rightarrow t_n,\Gamma}{\forexpr i {\tau} {b^d}}
    \\\\ \linresult
         {(\letx x {\forexpr i \tau e} \bullet)}
         {(\view j \tau {\ttt{fst}~x.j})}
         {(\forexpr j \tau {(\ttt{snd}~x.j~t_1 \ldots t_n)})}} \\
&\Infer{LinSlice}
  {}{\linrule{\diffenv}{v_1~\ttt{!}~v_2}
      \linresult{\bullet}{v_1~\ttt{!}~v_2}{\diffenv[v_1]~\ttt{!}~v_2}} \\
&\Infer{LinGet}
  {}{\linrule{\diffenv}{\ttt{get}~v_1~v_2}
     \linresult{\bullet}{\ttt{get}~v_1~v_2}{\ttt{get}~\diffenv[v_1]~\diffenv[v_2]}} \\
&\Infer{LinPut}
  {}{\linrule{\diffenv}{\ttt{put}~v_1~v_2}
      \linresult{\bullet}{\ttt{put}~v_1~v_2}{\ttt{put}~\diffenv[v_1]~\diffenv[v_2]}} \\
& \Infer{LinRunState}
  {\Delta = x_1\rightarrow t_1, \ldots x_n\rightarrow t_n,\Gamma \qquad
   \linreify{h\ann\ttt{Type},~x\ann\ttt{Ref}~h~\tau^d,~\Gamma}
       {h,~x,~x_1, \ldots, x_n}{b^d} e'}
  {\linrule{\diffenv}{\ttt{runState}~v^d~(\action h x {\ttt{Ref}~h~\tau^d} {b^d})} \\\\
    (\letx {((x_{\ttt{ans}}, x_{\ttt{lin}}), x_s)}
         {\ttt{runState}~v^d~(\action h x {\tau^d} e')} \bullet), \\\\
    (x_{\ttt{ans}}, x_s),\\\\
    \ttt{runState}~\diffenv[v^d]~(\action {h'} {x'} {\ttt{Ref}~h'~\tangentof\tau}(x_{\ttt{lin}}~h'~x'~t_1\ldots t_n))
  }
\end{align*}
\caption{Representative rules for linearization.  The linearization environment $\Delta$ carries tangent values, and also carries the primal type environment to be able to construct zero tangents when needed.}
\label{fig:linearize}
\end{figure}

\subsection{Transposition}
\label{sec:transposition}

In practice, AD is often used to compute the gradients of scalar-valued functions $f: \R^n \to \R$.
This is useful for gradient-based optimization in machine learning, or for sensitivity analysis of computational models to their parameters.

The \texttt{linearize} transform is semantically sufficient for this purpose, as given \texttt{f: (Fin n)=>Float -> Float}, one could compute the gradient by applying \texttt{snd (linearize f x)} to $n$ different inputs, each representing a standard basis vector of $\R^n$.
But the cost to compute the gradient would then be proportional to $n$ applications of \texttt{f}.
For neural networks with $n \simeq 10^8$, that's untenable!

The difficulty is that the result of linearization gives computational access to $\partial f(x)$ only through application, which corresponds to multiplying by the Jacobian only on the left.
Fortunately, structural linearity allows us to define another compile-time transformation that reverses the inputs and outputs of structurally linear functions:
\[
    \texttt{transpose : [VectorSpace a, VectorSpace b] (a -o b) -> (b -o a)}
\]
This operation is named \texttt{transpose} because it models transposition of a linear map.
%it's equivalent to transposing the Jacobian that its argument represents. \MATT{We don't need to refer to the matrix representation to define transposition, e.g. https://en.wikipedia.org/wiki/Transpose_of_a_linear_map.}

Computationally, \texttt{transpose} obeys similar desiderata to \texttt{linearize}:
\begin{itemize}
\item The computational cost of applying \texttt{transpose f} is within a small constant multiple of the cost of applying \texttt{f}.
\item The transposed function uses the same effects as the original \texttt{f}, except that repeatedly reading a value becomes associative accumulation with the (0, +) monoid, and vice versa.%
\footnote{The Dex effect system actually has a \texttt{Reader} effect to serve as the transpose of \texttt{Accum}, but a further compiler transform could in principle eliminate \texttt{Reader} and replace it with variable access.}
\item All structurally linear functions can be transposed, and the result is also structurally linear.
\end{itemize}

With \texttt{transpose}, we can compute gradients using
\begin{Verbatim}[xleftmargin=\codeindent,fontsize=\small]
grad : [VectorSpace a] (a -> Float) -> a -> a
grad f x = (transpose (snd (linearize f x))) 1.0
\end{Verbatim}
This recovers the desired effect of computing the gradient of \texttt{f} in time proportional to the runtime of \texttt{f}, and this is how Dex implements \emph{reverse-mode AD}.
See \citet{frostig2021decomposing} for further discussion of this approach.
When \ttt{transpose} is used to compute gradients this way, the intermediate values are conventionally called \emph{cotangents}.

Transposition is another source transform we implement in the Dex compiler, with a selection of rules displayed in~\Cref{fig:transpose}.
The
$$\transrule{\transenv}{e^d}{v} E$$
elaboration transposes the simplified structurally linear expression $e^d$ by accumulating into the references corresponding to its free linear variables,
starting with the cotangent value $v$, corresponding to the result of $e^d$. The environment $\transenv$ maps each variable to a reference
storing its (incrementally constructed) value in the transposed program.
% Each instance of {\ttfamily transpose (\textbackslash x:a. e)} is elaborated as:
% $$\verb|\y:b. snd (runAccum (\h:Type r:(Ref h a).|~\transruleop{x\rightarrow r}{e}{y}\verb|))|$$

Note that the rule \tsc{LetOtherTranspose} inverts the order of let-bindings in the block by sequencing $e_2$ to happen before $e_1$.
Similarly, the rule \tsc{ForTranspose} reverses the iteration order by substituting the index variable with its inverted counterpart.
This only matters if the body uses the state effect.

% The key feature of reverse-mode AD relevant to the sequel is visible in the type of \texttt{transpose}.
% % Namely, \texttt{transpose} must reverse the flow of data through its argument, computing from the output to the input.
% Recall that the argument to \texttt{transpose} is the linearization \texttt{lin: a -o b} of some primal function \texttt{f: a -> b}, which computes a directional derivative of \texttt{f} by propagating tangent values from \texttt{a} to \texttt{b}.
% Now, \texttt{transpose} promises to reverse the data flow of \texttt{lin}, and propagate derivative values (now conventionally called \emph{cotangents}) from \texttt{b} to \texttt{a}.  But, every non-linear primitive occurring in \texttt{f} needs to know its (primal) input to compute its derivative!
% Reverse-mode AD is thus a two-phase computation: first, the primal \texttt{f x} is evaluated in the forward direction, saving all necessary intermediate values; and then the cotangents are evaluated in the reverse direction, reading the primal intermediates as needed.
% In Dex, the forward phase is the partial evaluation performed by \texttt{linearize}, the residual from which stores the primal values.

We close this section by acknowledging that this decomposition of reverse-mode AD into forward-mode AD followed by transposition is unusual, and most AD systems just implement reverse-mode monolithically, sometimes not exposing forward-mode to users at all.
Unfortunately, formalizing this approach or defending its virtues would take us too far afield, but we hope the research community does that soon, at least more fully than \citet{frostig2021decomposing} did.

\begin{figure}\footnotesize
\begin{align*}
& \framebox{$\transrule{\transenv}{v^d}{v} e$} \\
& \Infer{VarTranspose}{}
     {\transrule {x\rightarrow r,~\transenv} {x} {t} {r~\ttt{+=}~t}}
 \qquad \Infer{ZeroTranspose}{}
     {\transrule {\transenv} {0.0} {t} {\bullet}} \\
& \Infer{PairTranspose}
     {      \transrule {\transenv} {v_1} {t_1} {e_1}
     \quad  \transrule {\transenv} {v_2} {t_2} {e_2}}
     {\transrule {\transenv} {(v_1,~v_2)} {t}
          {\letx {t_1} {\ttt{fst}~t}
           {\letx {t_2} {\ttt{snd}~t}
             (e_1;e_2)}}} \\
& \framebox{$\transrule{\transenv}{e^d}{v} e$} \\
& \Infer{AddTranspose}
     {     \transrule{\transenv}{v_1}{t} e_1
     \quad \transrule{\transenv}{v_2}{t} e_2}
     {\transrule{\transenv}{v_1~\ttt{+}~v_2}{t} {e_1}; {e_2}} \\
& \Infer{MulLeftTranspose}
     {     \transrule{\transenv}{v_1}{t'} e
     \quad \transenv \cap \freevars{v_2} = \emptyset }
     {\transrule{\transenv}{v_1~\ttt{*}~v_2}{t} \letx {t'} {t~\ttt{*}~v_2} e} \\
& \Infer{MulRightTranspose}
     {     \transenv \cap \freevars{v_1} = \emptyset
     \quad \transrule{\transenv}{v_2}{t'} e }
     {\transrule{\transenv}{v_1~\ttt{*}~v_2}{t} \letx {t'} {v_1~\ttt{*}~t} e} \\
& \Infer{ForTranspose}
  {\transrule{\transenv}{b^d}{t'}{e}} %%
  {\transrule{\transenv}{\forexpr i {\tau^d} {b^d}} {t}
      {{\forexpr i {\tau^d} \subst i {\ttt{reverse}~i}{(\letx {t'} {t.i} e)}}}} \\
& \Infer{GetTranspose}{}
  {\transrule{x\rightarrow v,~\transenv}{\ttt{get}~x}{t}
    \letx {t'} {{\ttt{get}~v}} {\ttt{put}~v~(t+t')}} \\
& \Infer{PutTranspose}{\transrule{\transenv}{v^d}{t'} e}
  {\transrule{x\rightarrow v',~\transenv}{\ttt{put}~x~v^d}{t}
    \letx {t'} {\ttt{get}~v'}
        {(\ttt{put}~v^d~\ttt{Zero}[\tau];~e)}}\\
& \Infer{RunStateTranspose}{
       \transrule{h\rightarrow h',~x\rightarrow x',~\transenv}
          {b^d}{t_{\ttt{ans}}} e_1   \qquad
       \transrule{\transenv}{v}{t_{\ttt{s}}'} e_2
    \qquad x',~h'~\mathrm{fresh}
  }
  {
    \transrule{\transenv}
    {\ttt{runState}~v~(\action {h} x {\ttt{Ref}~h~\tau^d} b^d)}
    {t} \\\\
      {\letx {(t_{\ttt{ans}},t_{\ttt{s}})} t
       \letx {((),t_{\ttt{s}}')}
           {\ttt{runState}~{t_{\ttt s}}~(\action {h'} {x'} {\ttt{Ref}~h'~\tau^d} e_1)}
           {} e_2}
  }\\
& \framebox{$\transrule{\transenv}{b^d}{v} e$} \\
& \Infer{LetNonlinearTranspose}
  {\Omega \cap \freevars{e^d} = \emptyset \qquad
   \transrule{\transenv} {b^d} t e'}
  {\transrule{\transenv}{\letx {x\ann{\tau^d}} {e^d} {b^d}}{t}
    \letx {x\ann{\tau^d}} {e^d} e'} \\
& \Infer{LetIndexingTranspose}
  {\transrule{x\rightarrow r', y\rightarrow r,~\transenv} {b^d} t e   }
  {\transrule{y\rightarrow r,~\transenv}{\letx {x\ann{\tau^d}} {y.v} {b^d}}{t}
    \letx {r'} {r!v} e} \\
& \Infer{LetSliceTranspose}
  {\transrule{x\rightarrow r', y\rightarrow r,~\transenv} {b^d} t e   }
  {\transrule{y\rightarrow r,~\transenv}{\letx {x\ann{\tau^d}} {y!v} {b^d}}{t}
    \letx {r'} {r!v} e} \\
& \Infer{LetOtherTranspose}
  {\Omega \cap \freevars{e^d} \neq \emptyset \qquad
   \transrule{x\rightarrow r,~\transenv} {b^d} {t} {e_2}  \qquad
   \transrule{\transenv} {e^d} {t'} {e_1}}
  {\transrule{\transenv}{\letx {x\ann{\tau^d}} {e^d} {b^d}}{t}
     (\letx {t'} {{(\ttt{runAccum}~\action h r {\ttt{Ref}~h~{\tau^d}} ~e_2)}} {e_1})} \\
\end{align*}
\vspace{-0.5cm}
\caption{A selection of transposition rules for the post-simplification language.}
\label{fig:transpose}
\vspace{-0.5cm}
\end{figure}

\subsection{Challenges posed by automatic differentiation}
\label{sec:ad-lessons}

We feel that Dex gained a great deal as a language from being co-designed with its automatic differentiation system.
AD is something like a very demanding user of the language---it is always trying to write programs the compiler developers did not anticipate, and always producing compelling bug reports or feature requests when those programs do not work or are slower than they should be.
In this section, we discuss a few specific subtleties in the design of Dex's AD, and the effects AD has had on the rest of the language and compiler.

\subsubsection{Capturing scoped intermediates}
\label{sec:capturing-scoped-intermediates}
% What if we need to tape e.g. intermediates that otherwise would never escape a \texttt{for} expression?
Non-linear primitive operations such as \texttt{mul} need to capture the intermediate values computed inside the differentiated function.
This is fine if the intermediate is in scope for the remainder of the function; but whenever we linearize an expression that may construct intermediates that go out of scope (such as \texttt{for}), we have to arrange for their values to be captured.

This is why we define differentiation to return core IR rather than the post-simplification subset of it.
By emitting core IR, \ttt{linearize} and \ttt{transpose} can just capture whatever they need in a fresh lambda expression, whose type then doesn't need to reflect the type of the data it is carrying in the closure.
We then recover post-simplification IR by running simplification again on the output.
This architecture does mean that simplification and differentiation have to form a loop in the compiler, rather than being sequential passes, but we feel that the simplification of linearization and transposition thus won is worth it.

\subsubsection{Differentiation of state-mutating code}
\label{sec:ad-through-state}
% What happens if the differentiable operators need a piece of state that is overwritten later?
Reverse-mode automatic differentiation is notoriously difficult in the presence of mutation in the program being differentiated.
The issue is again storage of intermediate primal values.
But what if one of these primals is mutated by a later operation in the program?

In Dex, this problem disappears, because the type system distinguishes between the (mutable) \emph{reference} being read and/or updated, and the (immutable) \emph{value} one obtains from it with \verb|get|.
The~value is saved by \texttt{linearize} as needed, by the same mechanism as all other values.
The data in the reference itself does not need to be stored, because \verb|get| is linear: knowing the cotangent of the value emitted is enough to compute the update that must be made to the mutable buffer, and no additional information about the primal data therein is needed.
Conversely, the transpose of \verb|put| just reads the currently accumulated cotangent of the reference.

Even in pure code, it remains important that only \emph{non-linear} operations store their inputs.
Many Dex loops operate on just one or a few index values of the state array per iteration, so
making a complete copy of the state at each step could raise even the asymptotic complexity of the linearized function.
Array indexing, however, is linear, so all those copies can be avoided.
Indeed, given an expression such as \texttt{for i. sin (get ref).i}, only the argument of \ttt{sin} needs to be saved for each iteration of the loop.

\subsubsection{Transposing array indexing}
\label{sec:ad-of-indexing}
The cotangent of array indexing is a recurring problem in the design of automatic differentiation systems.
The reverse-mode update due to reading \texttt{x = array.i} is of course the sparse update \verb|array_cot!i += x_cot|, but how should we represent this?

In an imperative language, one is always free to just emit the mutating update, which has been the de-facto standard implementation technique of AD systems over decades.
This correctly conserves work, but turns what used to be a parallelizable loop reading \texttt{array} into a sequential loop writing to \verb|array_cot|.
Of course, the parallelism can be recovered by downstream compiler analyses in simple cases, but complex cases are not difficult to come by.

On the other hand, in pure array languages (such as JAX \cite{jax2018github} and other Python-based array-libraries) the conventional pattern used to implement reverse-mode AD is to just one-hot encode the indexing update, i.e., create an update array of almost all zeros.
This is simple to implement, and costs no parallelism, but of course creates an asymptotic increase in memory usage and count of arithmetic operations.
Given the importance of indexing in Dex, this is not a viable option for us.

Another alternative in a pure system is to perform cotangent accumulation functionally by accumulating the updates in sparse data structures that gather the index or indices where an update is to occur along with the value(s) to be added there.
This is also asymptotically work-preserving, but imposes large constant or logarithmic factor overhead in performance, as well as a significant constant factor in developer effort.
It is also quite unfriendly to hardware acceleration due to the inherent dynamism in the shape of the data structure and requirements on dynamic memory allocation.
We are thus not aware of any AD system that actually differentiates indexing this way.

Instead, we note that while general mutation is certainly sufficient to preserve work, it's not actually necessary---the only mutation that reverse-mode AD needs is to accumulate sums.
More importantly, because summation is associative (up to floating-point rounding), it can still be executed in parallel.
The desire to capture this was the prime reason why we introduced an effect system into Dex, and in particular why we defined \texttt{Accum}.
The \texttt{Accum} effect is expressive enough to capture cotangent updates, but restrictive enough to be easy to parallelize (Section~\ref{sec:auto-parallel}).
Having introduced \texttt{Accum} into the language, we promptly found other uses for it besides cotangent accumulation, for instance the \texttt{histogram} example from~\Cref{sec:work-efficiency}.

% \todo[inline]{Dougal explains or references how Accum is differentiated, preserving closure.}
% \todo[inline]{Example?}
% TODO: Talk about transposition of loops?
% \subsubsection{Parallelism} Reverse-mode inverts data dependencies, making it difficult to preserve parallelism

\section{Performance and parallelism}
\label{sec:performance}

In this section we describe the remainder of the Dex compiler.  After simplification (\Cref{sec:defunctionalization}) and differentiation (\Cref{sec:ad}):
\begin{enumerate}
\item Dex optimizes the post-simplification and differentiated IR with standard techniques: inlining, dead code elimination, common subexpression elimination, loop invariant code motion, etc. We comment in \Cref{sec:fusion} on why Dex does not need a suite of operation fusion optimizations.
\item Dex generates LLVM bytecode for final compilation for CPU or GPU.
Code generation is target-aware: Dex chooses different LLVM instructions depending on the hardware being compiled for.
The interesting part is mapping Dex structures to parallel execution strategies, which we cover in \Cref{sec:auto-parallel}.
\end{enumerate}
Finally, in \Cref{sec:type-directed-compilation} we comment on a few places where Dex's rich type system makes the compiler simpler and more effective, and present some preliminary benchmark results in \Cref{sec:benchmarks}.

\subsection{Automatic parallelization}
\label{sec:auto-parallel}

The surface language of Dex does not expose any way for the user to directly express an intent to evaluate a number of expressions in parallel.
Instead, all parallelism-related decisions are made by~the~compiler.
In this section, we describe Dex's automatic parallelization algorithm (\Cref{sec:parallelism-traversal}), and then comment on how Dex's \verb|Accum| effect navigates a common work-parallelism tradeoff (\Cref{sec:work-efficiency}) and allows the programmer to smoothly shift among different-seeming parallel computation patterns (\Cref{sec:parallel-accum}).

\subsubsection{Parallelism allocation}
\label{sec:parallelism-traversal}
Many modern hardware accelerators are highly constrained, and can only efficiently parallelize programs that transform batches of data in a uniform manner (much like the SIMD units in CPUs).
Dex makes parallelization decisions at the granularity of \texttt{for} expressions, because they have exactly this characteristic: a sequence of instructions (the body expression) is evaluated repeatedly over slightly different data (the loop indices and values derived from them).

Furthermore, accelerators generally do not support nested parallelism, meaning that the total number of parallel invocations has to be decided at the top level and cannot be increased later.
Due to this, Dex flattens nested \texttt{for} expressions that can be run in parallel.
We do not describe this procedure formally, because it is largely analogous to the one proposed for Futhark by \citet{henriksen2017futhark} (if only one replaces Futhark's \texttt{map} with \texttt{for}, while treating the \texttt{Accum} effect similarly to \texttt{reduce} and \texttt{State} to \texttt{loop}).

\subsubsection{Work-efficiency guarantees}
\label{sec:work-efficiency}

While we approach parallelism extraction highly analogously to Futhark, it is a much less critical optimization for Dex than it is for array-combinator languages.
The reason is that there are programs that can be expressed in most array-combinator languages in only two ways: one that is work-efficient but fully sequential, and another that is parallelizable but is \emph{not} work-efficient.
One good example of such a program is histogram calculation:
\begin{Verbatim}[xleftmargin=\codeindent,fontsize=\small]
histogram_seq = \points:(n=>k).
  yieldState (for i:k. 0) \hist.
    for i. hist!(points.i) += 1

histogram_par = \points:(n=>k).
  sum (map one_hot points)
one_hot = \idx:k. for i. if i == idx then 1 else 0
\end{Verbatim}
The \verb|histogram_seq| function has the right run-time complexity of $O(n + k)$ (where $n$ and $k$ are the sizes of the index sets \verb|n| and \verb|k|), but uses a stateful loop and so it cannot be made parallel without sophisticated analysis.
On the other hand, \verb|histogram_par| is a composition of the two parallel operators \texttt{map} and \texttt{sum}, but a naive sequential execution would suffer the significantly worse asymptotic complexity of $O(nk)$.
Since the goal of array-combinator languages is to enable parallel execution, the first approach is considered undesirable.
Instead, array-combinator languages \emph{have to} perform an optimization step that turns the abundant parallelism present in \verb|histogram_par| into a partially sequential loop, that lets them eventually achieve the desired work efficiency (see~\citet{henriksen2017futhark} for details) or, as many do, accept the sequential performance penalty.

Meanwhile, in Dex, the same program can be expressed with an associative accumulator effect:
\begin{Verbatim}[xleftmargin=\codeindent,fontsize=\small]
histogram_dex = \points:(n=>k).
  yieldAccum \hist.
    for i. hist!(points.i) += 1
\end{Verbatim}
If the \texttt{for} expression's body was to be evaluated sequentially, it would have the right complexity of $O(n + k)$.
But, because the \texttt{Accum} effect exposes enough structure, the Dex compiler is able to take this work-efficient implementation and evaluate the body in parallel too.
%, as outlined in Section~\ref{sec:parallelism-traversal}.

In short, while we expect the eventual evaluation strategy employed by both approaches to be largely equivalent, Dex has the benefit of being able to naturally express a parallel \emph{and} work-efficient program instead of relying on opaque compiler optimizations.
While it might seem like a minor point, we strongly believe that this is a big step forward in terms of usability.
Compiler optimizations are certainly useful, but they are largely outside of the control of the user and as such they should not be the only way to achieve the right asymptotics.
% \apc{TODO: Have a catchy phrase about how optimizations should be optional, and even programs without optimizations should at least have the right run-time complexity. Combinator languages fail to guarantee that.}

\subsubsection{Reduction patterns by indexing}
\label{sec:parallel-accum}

Every use of the accumulation effect corresponds to some form of reduction, but it is useful to classify them by strategies that can carry out different reductions in parallel.
The simplest pattern is a \emph{complete reduction}, where a large set of values is reduced to just a single one (and we do not try to take advantage of any structure the result might have).
Then, a \emph{regular segmented reduction} occurs when the reduced value is an array of equal-size segments, each of which contributes one component of the result.
In that case, each parallel thread can be assigned a subset of segments to reduce and will not need to replicate the full accumulator, but only the entries it computes.
Multiple compilation strategies for this pattern have been implemented in Futhark \cite{larsen2017strategies}.
Finally, one gets an \emph{irregular segmented reduction} or a \emph{histogram} when the different segments are of varying sizes that may be difficult to predict even at run-time (for example, because the data of different segments is interleaved).
Yet again, a special routine for this case has been implemented recently in Futhark \cite{henriksen2020histograms}.

In an array-combinator language, these three types of reductions are naturally expressed as distinct combinators.
In Dex, however, they are just slightly different indexing patterns for the accumulator reference:
\begin{itemize}
\item A complete reduction is when the accumulator is never indexed.
\item A regular segmented reduction is when the accumulator is indexed by a subset of the loop indices.
\item An irregular segmented reduction is when the indices of the accumulator are derived form arbitrary expressions (such as an array lookup).
\end{itemize}
Those differences are readily visible in the following example:
\begin{Verbatim}[xleftmargin=\codeindent,fontsize=\small]
complete_reduction = \values:n=>Float.
  yieldAccum \acc. for i. acc += values.i

segmented_reduction = \values:n=>m=>Float.
  yieldAccum \acc. for i j. acc!i += values.i.j

histogram = \classes:n=>k.
  yieldAccum \hist. for i. hist!(points.i) += 1
\end{Verbatim}

While Dex does not take advantage of this observation and treats all reductions as though they were complete, we can imagine recovering more sophisticated execution strategies with a more carefully effect-aware loop parallelization transform.

\subsection{Fusion optimizations}
\label{sec:fusion}

One advantage of Dex's program representation is that Dex does not need a large and complex suite of fusion rules to achieve good performance.

Indeed, one of the drawbacks of the bulk array programming model is that each individual bulk operation needs to encapsulate enough work to amortize away the overhead of dispatching a kernel for it and allocating storage for its inputs and outputs.
Expressiveness, however, calls for composing programs out of many small operations, because they can be put together in many different ways.
Traditional array-combinator languages bridge this gap with \emph{fusion} optimizations, each of which combines a pattern of smaller operations into a larger one.
The quintessential example is combining a sequence of two \texttt{map} operations into one \texttt{map} of the composition of the two functions, thus eliding the intermediate array.

Unfortunately, such fusion rules have to be specified for almost every pair of combinators, leading to a drastic increase in the complexity of the system.
To make matters even worse, fusion might necessitate adding more combinators to capture computation patterns which the fusion rules can emit; and these new combinators then need more fusion rules of their own.
For example, Futhark uses an otherwise-redundant combinator named \texttt{redomap} \citep{henriksen2016design} to represent a computation that constructs an array while reducing a set of values.
% This combinator adds nothing to Futhark's expressiveness, since it's just a composition of \texttt{map} and \texttt{reduce}, but is necessary for closure under Futhark's fusion rules.

In Dex, in contrast, all fusion rules can be seen as instances of inlining followed by reduction.
For example, consider this block, showing a reduction happening along a map, with its result being consumed in another map:
\begin{Verbatim}[xleftmargin=\codeindent,fontsize=\small]
y = for j.
      acc += f j
      g j
x = for i. h y.i
\end{Verbatim}
\mbox{}\vspace{-0.5cm} \\
Assuming that this is the only use of the value \texttt{y}, each element of \texttt{y} is consumed exactly once.
This means that we can safely inline the \texttt{for} expression that constructs it, for as long as we reduce the \texttt{(for j. ...).i} form to avoid duplicating effects.
This yields just a single loop that avoids materializing the \texttt{y} array, but produces each of its elements on demand instead:
\begin{Verbatim}[xleftmargin=\codeindent,fontsize=\small]
x = for i.
      acc += f i
      h (g i)
\end{Verbatim}

Replicating a high-quality set of loop fusion optimizations in Dex is just a matter of doing a good job of inlining \texttt{for} expressions.
Fortunately, the literature on inlining strategies is vast \cite{peytonjones2002secrets, mitchell2010rethinking}.
This is also where the effect system comes in handy, as it lets us easily decide which inlining decisions are legal (i.e., do not duplicate nor reorder effects).
%\todo[inline]{Cite some stuff on using effects to help determine legality of optimization moves.}

\subsection{Type-directed compilation}
\label{sec:type-directed-compilation}

Dex's type information is also useful for making optimization decisions and emitting efficient code. In addition to the usual benefits of type-directed compilation,
explicit index set types ensure that no out-of-bounds access could ever happen upon array indexing---a type-safe index value cannot, by definition, be out of bounds.
In general, a run-time bounds check is still necessary when the index is constructed, e.g., by \texttt{fromOrdinal}.
However, the vast majority of indices are constructed by \texttt{for} expressions, and in that case the bounds checks can be trivially elided, since we know that the ordinals a \texttt{for} iterates over are always in bounds.

Moreover, having the full (nested) array type available at lowering time, we are able to perform an array-of-structures to structure-of-arrays layout conversion which usually achieves much better performance on parallel architectures.
After this conversion, we end up with a number of nested array types that only contain primitive scalar types, which makes it possible for us to emit code that computes the total number of array elements (or in many cases even derive it statically), and allocate a single flat memory buffer that holds unboxed values.
Ultimately, this means that the rich array abstraction always gets translated into just a few pointers to unboxed data, along with a few integers used to compute offsets into the flattened arrays.
As a result, Dex programs generally should not incur any significant overheads compared to the languages traditionally used for low-level array computing such as Fortran, C, and C++.

As mentioned previously, having explicit effect types allows us to statically distinguish the loops that are embarassingly parallel (no effects), parallelizable with some care (only a number of \texttt{Accum} effects), or necessarily serial (when the loop body induces the \texttt{State} effect).
The fine-grained effect system presents us with both a simple compiler pipeline, and a straightforward mental framework that allows our users to understand the parallelization opportunities their code exposes.

\ExcessDetail{
Finally, type-checking the IR after every transformation has been invaluable for debugging the compiler.
}

\subsection{Evaluation}
\label{sec:benchmarks}

As an early test of the effectiveness of our language and compilation strategy, we compare the run-time of Dex programs against Futhark \cite{henriksen2017futhark} on four programs from two benchmark suites.
We chose Futhark as a baseline because it is one of the closest languages to Dex in purpose and structure, and because it has already been shown to have broadly competitive performance with the low-level (usually C++/CUDA based) implementations by the original benchmark authors.

One important caveat: the benchmark programs we have selected are not necessarily representative of common array workloads.
Specifically, given the relative unsophistication of our optimization pipeline, we do not expect Dex to be competitive yet on workloads that depend critically on specific, heavily hand-optimized compute-intensive kernels like matrix mutiplication.
Instead, we selected the benchmark problems to highlight tasks that stress compilation of more-general array programs, where eliminating composition overheads is important.
% The goal of this paper is to present a fresh way of looking at numerical programming without giving up on performance, not a way to achieve speedups over potentially more constrained DSLs.

\subsubsection{Benchmark results}
\label{sec:benchmark-results}
% gcloud beta compute --project=<your project> instances create dex-bench --zone=us-central1-a --machine-type=n1-standard-8 --subnet=default --network-tier=PREMIUM --maintenance-policy=TERMINATE --no-service-account --no-scopes --accelerator=type=nvidia-tesla-v100,count=1 --min-cpu-platform="Intel Skylake" --image=ubuntu-2004-focal-v20210223 --image-project=ubuntu-os-cloud --boot-disk-size=256GB --boot-disk-type=pd-ssd --boot-disk-device-name=dex-bench --no-shielded-secure-boot --shielded-vtpm --shielded-integrity-monitoring --reservation-affinity=any

% wget https://developer.download.nvidia.com/compute/cuda/repos/ubuntu2004/x86_64/cuda-ubuntu2004.pin
% sudo mv cuda-ubuntu2004.pin /etc/apt/preferences.d/cuda-repository-pin-600
% sudo apt-key adv --fetch-keys https://developer.download.nvidia.com/compute/cuda/repos/ubuntu2004/x86_64/7fa2af80.pub
% sudo add-apt-repository "deb https://developer.download.nvidia.com/compute/cuda/repos/ubuntu2004/x86_64/ /"
% sudo apt-get update
% sudo apt-get -y install cuda

% sudo apt-get install -y haskell-stack llvm-9-dev clang-9 libpng-dev
% stack upgrade
% git clone https://github.com/google-research/dex-lang
% cd dex-lang
% TODO: checkout a particular commit?
% make

We compare Dex to Futhark on four problems:
\begin{itemize}
\item Hotspot, a stencil computation that solves heat equations from Rodinia \cite{che2009rodinia};
\item Pathfinder, a dynamic program, also from Rodinia;
\item MRI-Q, a standard map-reduce operation from the Parboil suite \cite{stratton2012parboil}; and
\item Stencil, a 3-D stencil computation, also from Parboil.
\end{itemize}
Three of these benchmarks define multiple data sizes on which they can be evaluated, and we include this breakdown as well.

\begin{figure}[b]
\centering
\begin{subfigure}{.5\textwidth}
  \centering
  \includegraphics[width=\textwidth]{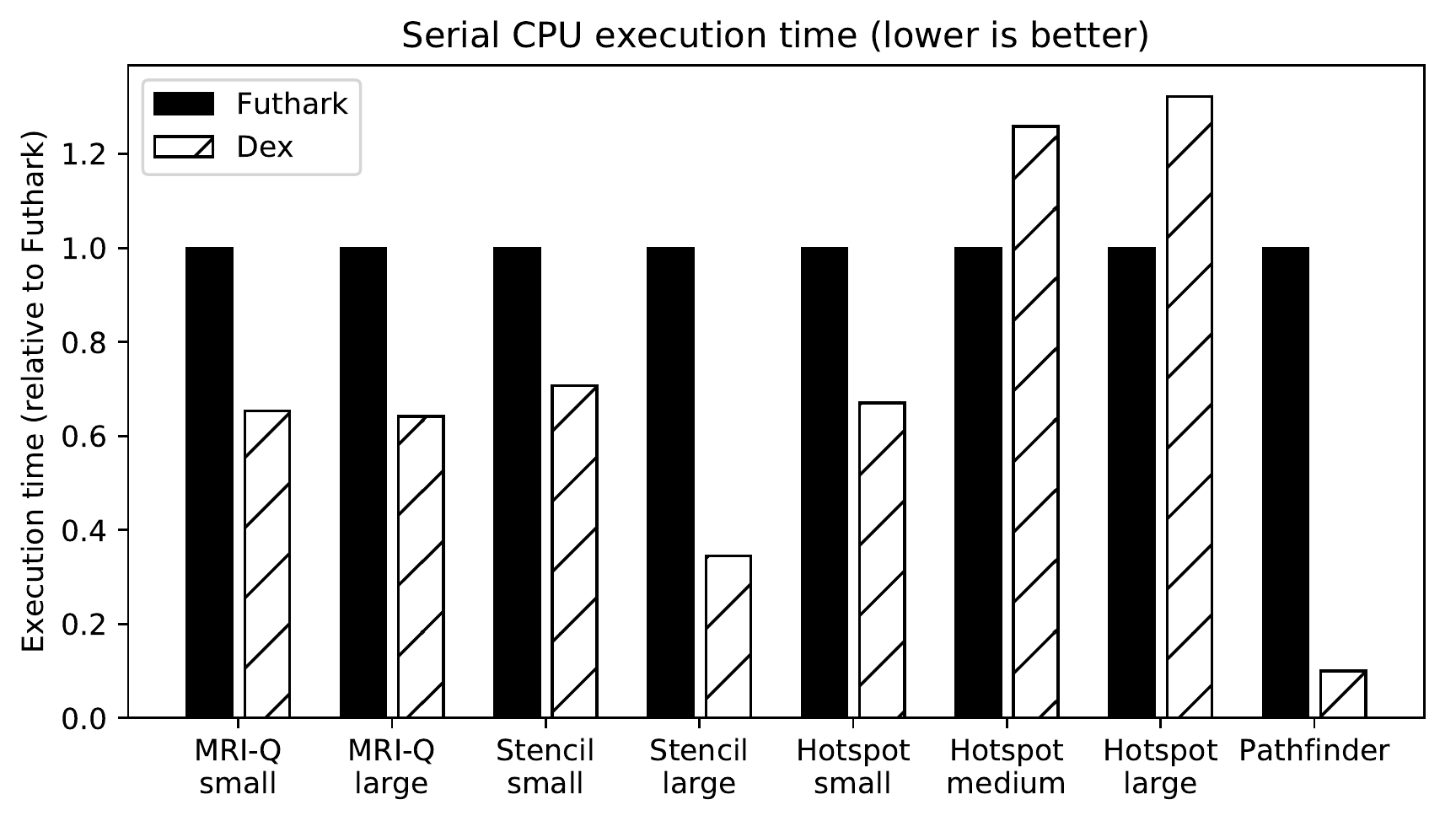}
  \caption{Serial CPU results}
  \label{fig:cpu-benchmarks}
\end{subfigure}%
\begin{subfigure}{.5\textwidth}
  \centering
  \includegraphics[width=\textwidth]{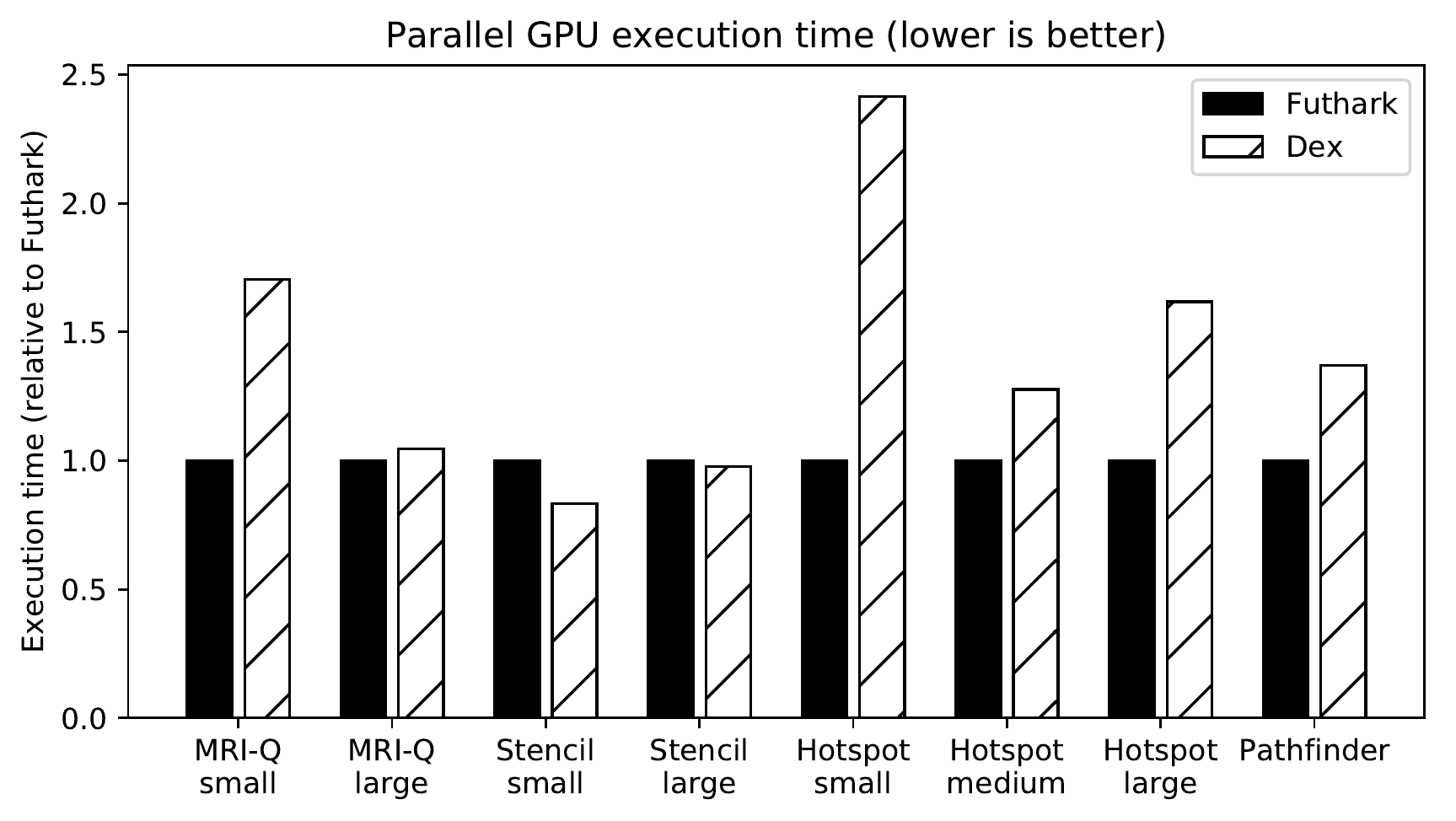}
  \caption{Parallel GPU results}
  \label{fig:gpu-benchmarks}
\end{subfigure}
\caption{Comparison of Dex and Futhark execution times in serial and parallel settings. Note that all times are normalized and relative to the run time of the respective Futhark programs. The Y axis of both figures grows with execution time, so less is better. The Futhark programs were authored by the Futhark developers and \href{https://github.com/diku-dk/futhark-benchmarks}{published on GitHub} for evaluation against their respective benchmark suites; we recompiled and re-ran them on the same hardware as our Dex programs for those benchmarks.}
\label{fig:benchmarks}
\end{figure}

The results can be seen in \Cref{fig:benchmarks}.
To aid reproducibility, we have used an \texttt{n1-standard-8} Google Cloud instance (30GB RAM, 8 vCPUs) with a single NVIDIA V100 GPU and an Intel Skylake CPU.
The relevant software versions are: CUDA 11.2, Dex from commit \texttt{069781e} and Futhark 0.18.6.

The general trend is that Dex is somewhat faster than Futhark on serial CPU execution, and somewhat slower on parallel GPU execution.
The worst slowdowns in the parallel setting can be observed on the smallest benchmarks, which are small enough so that the overheads associated with calling the CUDA functions become noticeable.
The difference usually becomes less pronounced as the size of the data grows, which is the most important use case for parallel accelerators.

We emphasize that our goal is not to demonstrate that Dex delivers performance improvements over existing systems.
Rather, the goal at this stage is to check that Dex can more or less match the~state of the art, while exploring a novel approach to expressing numerical computing workloads.
As Dex matures and we devote more effort to its optimization pipeline, we expect its absolute performance to improve, though the same is of course true of comparable systems as well.

% \subsubsection{AD performance}

% \apc{TODO: What should we compare to? Tapenade? What workloads? AD bench from MSFT?}

\section{Related work}
% We should at least acknowledge that, and possibly actually make the argument.
\label{sec:related-work}

While many ideas present in Dex have been explored separately, the language can be seen as a synthesis of a wide range of long-standing research topics.

\subsection{Array languages}
\label{sec:related-array-languages}
Dex follows a long line of array-oriented and data-parallel languages, dating back to APL \cite{iverson1962apl} and NESL \cite{blelloch1993nesl}, which has recently seen a surge in activity \cite{peytonjones2008parallelhaskell,ragankelley2013halide,mcdonell2013optimizing,slepak2014remora,steuwer2017lift,shaikhha2019fsmooth,hu2019taichi}.
The most widely-used today are first-order nd-array languages in the NumPy family \cite{harris2020numpy,jax2018github,paszke2019pytorch,abadi2016tensorflow}, whose advantages and disadvantages are discussed in \Cref{sec:introduction}.
Another relevant body of related work is in programming languages based on array combinators, treated in \Cref{sec:array-combinator-languages}.
Futhark is arguably the most similar language to Dex \cite{henriksen2017futhark}.
Similarly to the typed index sets in Dex, it tracks and propagates array shapes in its type system, making it possible to catch a whole class of (potential) shape errors at compile time.
In-scope variables of integer type can be implicitly lifted to bounds for array dimensions, leading to a mechanism similar to the \texttt{Fin} index set constructor in Dex.
Futhark also supports stateful computations, although it does so via a uniqueness typing system that guarantees safe in-place modification, instead of using an effect system like we do.

\subsection{Type systems}
\label{sec:type-systems}

The Dex type system is based on a number of previously published ideas.
While it breaks the highly conventional value-type boundary, it does so only by introducing a limited form of dependent typing, based on value-dependent types \cite{swamy2011valuedependent}.
In particular, types cannot depend on arbitrary expressions, but only on fully-evaluated values.
Dex's effect system is modeled on the row-polymorphic effect system of Koka \cite{leijen2014koka} and mostly implements a subset of it, with the slight exception of extending it with the reference slicing operation.
However, to the best of our knowledge, the use of an associative state effect (\texttt{Accum}) in order to emit parallel code has not been considered previously in any other effect system.
Our type inference algorithm largely follows the ideas of bidirectional type inference outlined by \citet{jones2007practicaltypeinference}, although in Dex higher-rank types are generally replaced by functions with implicit arguments.

% \subsection{Effect systems for safe parallelism}

% Delite and Lime?

% \subsection{Imperative languages}

% \paragraph{Julia} ...

% \paragraph{Chapel \cite{chamberlain2020chapel}} ...

% \paragraph{ispc} ...

\subsection{Automatic differentiation}
\label{sec:related-ad}

Automatic differentiation has a deep history \citep{baydin2017adsurvey,griewank2008evaluating,pearlmutter2008lambda}.
Dex benefits from many influences, especially \citet{elliott2018ad} and \citet{jax2018github}.
For brevity, we focus the discussion on array- and parallelism-oriented AD systems.
For a more complete treatment, see the survey \citet{baydin2017adsurvey} and the related works of \citet{shaikhha2019fsmooth} and \citet{bernstein2020differentiating}.

Machine learning has made great use of automatic differentiation.
Popular systems like Theano \citep{bergstra2010theano}, Autograd \citep{maclaurin2015autograd}, Chainer \citep{tokui2019chainer}, TensorFlow \citep{abadi2016tensorflow}, DiffSharp \citep{baydin2017adsurvey}, PyTorch \citep{paszke2019pytorch}, and JAX \citep{jax2018github} employ the bulk array programming model, relying on accelerator-friendly parallelism within each bulk operation.
This model is great for AD, since linearization and transposition mostly preserve program structure and hence performance.
But array indexing is a weak spot, especially in the context of loops: transposing a loop of indexed reads falls outside the set of performant programs, resulting in programs that use too much memory, too many FLOPs, or too little parallelism.
There are mitigation techniques; for example, Autograd and TensorFlow add a runtime sparse data representation, while JAX relies on downstream whole-program optimizations.
But these can be brittle, add overheads, or yield the wrong asymptotic complexity.
For the most part, these AD systems work well because users constrain themselves to using bulk operations rather than more flexible loops and indexing.

Systems that support performant loops along with indexed reads and writes, like ADIFOR \citep{bischof1992adifor}, Tapenade \citep{hascoet2013tapenade}, Zygote.jl \citep{Zygote.jl-2018}, RelayIR \citep{roesch2018relay}, and DiffTaichi \citep{hu2020difftaichi}, are better equipped to provide AD with the right asymptotic complexity.
But indexed writes, introduced by transposition of indexed reads, typically must be sequenced, giving up parallelism and accelerator-friendliness.
The aim of Dex's AD is to support these performant loops with indexing, yet preserve parallelism through typed effects.

ATL \citep{bernstein2020differentiating} also tackles this issue head-on, achieving AD of loopy indexing code while preserving both complexity and parallelism.
The core approach is to encode sparsity with APL's Iverson bracket, then rely on compiler optimizations.
The language is carefully designed along with these optimizations so that the system is provably efficient, rather than handling only important special cases \citep{huckelheim2019automatic,li2018halidead}.
This is powerful, but limits expressiveness.
Overall, guaranteed sparsity optimizations like ATL's and explicit \texttt{Accum} effects like Dex's may prove complementary.

\section{Conclusion}
We presented a synthesis of ideas from functional programming intended to support stateful and ``pointful'' numerical code in a richly-typed functional language.
From the user's point of view, we expect that this approach can support a programming style roughly as expressive and flexible as low-level imperative numerical code, while catching a wider range of bugs at compile time and being more concise through high-level abstractions.
From the compiler's point of view, a fine-grained typed effects system makes it possible to implement optimizations and program transformations (such as automatic differentiation and structurally-linear transposition) in a way that robustly preserves performance.

\begin{acks}

We would like to thank Dan Zheng, Sasha Rush and Lyndon White for their contributions to the open source implementation of Dex.
We are also very grateful for many helpful conversations with Roy Frostig, Gilbert Bernstein, George Necula, Martin Abadi and Gordon Plotkin.

\end{acks}

\bibliographystyle{ACM-Reference-Format}
\bibliography{main}


\begin{thebibliography}{52}

%%% the Web address.

\ifx \showCODEN    \undefined \def \showCODEN     #1{\unskip}     \fi
\ifx \showDOI      \undefined \def \showDOI       #1{#1}\fi
\ifx \showISBNx    \undefined \def \showISBNx     #1{\unskip}     \fi
\ifx \showISBNxiii \undefined \def \showISBNxiii  #1{\unskip}     \fi
\ifx \showISSN     \undefined \def \showISSN      #1{\unskip}     \fi
\ifx \showLCCN     \undefined \def \showLCCN      #1{\unskip}     \fi
\ifx \shownote     \undefined \def \shownote      #1{#1}          \fi
\ifx \showarticletitle \undefined \def \showarticletitle #1{#1}   \fi
\ifx \showURL      \undefined \def \showURL       {\relax}        \fi
% invisible to TeX
\providecommand\bibfield[2]{#2}
\providecommand\bibinfo[2]{#2}
\providecommand\natexlab[1]{#1}
\providecommand\showeprint[2][]{arXiv:#2}

\bibitem[\protect\citeauthoryear{Abadi, Barham, Chen, Chen, Davis, Dean, Devin,
  Ghemawat, Irving, Isard, Kudlur, Levenberg, Monga, Moore, Murray, Steiner,
  Tucker, Vasudevan, Warden, Wicke, Yu, and Zheng}{Abadi et~al\mbox{.}}{2016}]%
        {abadi2016tensorflow}
\bibfield{author}{\bibinfo{person}{Mart\'{\i}n Abadi}, \bibinfo{person}{Paul
  Barham}, \bibinfo{person}{Jianmin Chen}, \bibinfo{person}{Zhifeng Chen},
  \bibinfo{person}{Andy Davis}, \bibinfo{person}{Jeffrey Dean},
  \bibinfo{person}{Matthieu Devin}, \bibinfo{person}{Sanjay Ghemawat},
  \bibinfo{person}{Geoffrey Irving}, \bibinfo{person}{Michael Isard},
  \bibinfo{person}{Manjunath Kudlur}, \bibinfo{person}{Josh Levenberg},
  \bibinfo{person}{Rajat Monga}, \bibinfo{person}{Sherry Moore},
  \bibinfo{person}{Derek~G. Murray}, \bibinfo{person}{Benoit Steiner},
  \bibinfo{person}{Paul Tucker}, \bibinfo{person}{Vijay Vasudevan},
  \bibinfo{person}{Pete Warden}, \bibinfo{person}{Martin Wicke},
  \bibinfo{person}{Yuan Yu}, {and} \bibinfo{person}{Xiaoqiang Zheng}.}
  \bibinfo{year}{2016}\natexlab{}.
\newblock \showarticletitle{TensorFlow: A System for Large-Scale Machine
  Learning}. In \bibinfo{booktitle}{\emph{Proceedings of the 12th USENIX
  Conference on Operating Systems Design and Implementation}}
  \emph{(\bibinfo{series}{OSDI'16})}. \bibinfo{publisher}{USENIX Association},
  \bibinfo{address}{USA}, \bibinfo{pages}{265–283}.
\newblock
\showISBNx{9781931971331}


\bibitem[\protect\citeauthoryear{Baydin, Pearlmutter, Radul, and
  Siskind}{Baydin et~al\mbox{.}}{2017}]%
        {baydin2017adsurvey}
\bibfield{author}{\bibinfo{person}{At\i{}l\i{}m~G\"{u}nes Baydin},
  \bibinfo{person}{Barak~A. Pearlmutter}, \bibinfo{person}{Alexey~Andreyevich
  Radul}, {and} \bibinfo{person}{Jeffrey~Mark Siskind}.}
  \bibinfo{year}{2017}\natexlab{}.
\newblock \showarticletitle{Automatic Differentiation in Machine Learning: A
  Survey}.
\newblock \bibinfo{journal}{\emph{J. Mach. Learn. Res.}} \bibinfo{volume}{18},
  \bibinfo{number}{1} (\bibinfo{date}{Jan.} \bibinfo{year}{2017}),
  \bibinfo{pages}{5595–5637}.
\newblock
\showISSN{1532-4435}


\bibitem[\protect\citeauthoryear{Bergstra, Breuleux, Bastien, Lamblin, Pascanu,
  Desjardins, Turian, Warde-Farley, and Bengio}{Bergstra et~al\mbox{.}}{2010}]%
        {bergstra2010theano}
\bibfield{author}{\bibinfo{person}{James Bergstra}, \bibinfo{person}{Olivier
  Breuleux}, \bibinfo{person}{Fr{\'e}d{\'e}ric Bastien},
  \bibinfo{person}{Pascal Lamblin}, \bibinfo{person}{Razvan Pascanu},
  \bibinfo{person}{Guillaume Desjardins}, \bibinfo{person}{Joseph Turian},
  \bibinfo{person}{David Warde-Farley}, {and} \bibinfo{person}{Yoshua Bengio}.}
  \bibinfo{year}{2010}\natexlab{}.
\newblock \showarticletitle{Theano: a {CPU} and {GPU} math expression
  compiler}. In \bibinfo{booktitle}{\emph{Proceedings of the Python for
  scientific computing conference (SciPy)}}, Vol.~\bibinfo{volume}{4}. Austin,
  TX, \bibinfo{pages}{1--7}.
\newblock


\bibitem[\protect\citeauthoryear{Bernstein, Mara, Li, Maclaurin, and
  Ragan-Kelley}{Bernstein et~al\mbox{.}}{2020}]%
        {bernstein2020differentiating}
\bibfield{author}{\bibinfo{person}{Gilbert Bernstein}, \bibinfo{person}{Michael
  Mara}, \bibinfo{person}{Tzu-Mao Li}, \bibinfo{person}{Dougal Maclaurin},
  {and} \bibinfo{person}{Jonathan Ragan-Kelley}.}
  \bibinfo{year}{2020}\natexlab{}.
\newblock \showarticletitle{Differentiating a Tensor Language}.
\newblock \bibinfo{journal}{\emph{arXiv preprint arXiv:2008.11256}}
  (\bibinfo{year}{2020}).
\newblock


\bibitem[\protect\citeauthoryear{Bischof, Carle, Corliss, Griewank, and
  Hovland}{Bischof et~al\mbox{.}}{1992}]%
        {bischof1992adifor}
\bibfield{author}{\bibinfo{person}{Christian Bischof}, \bibinfo{person}{Alan
  Carle}, \bibinfo{person}{George Corliss}, \bibinfo{person}{Andreas Griewank},
  {and} \bibinfo{person}{Paul Hovland}.} \bibinfo{year}{1992}\natexlab{}.
\newblock \showarticletitle{{ADIFOR} --- generating derivative codes from
  {F}ortran programs}.
\newblock \bibinfo{journal}{\emph{Scientific Programming}} \bibinfo{volume}{1},
  \bibinfo{number}{1} (\bibinfo{year}{1992}), \bibinfo{pages}{11--29}.
\newblock


\bibitem[\protect\citeauthoryear{Blelloch}{Blelloch}{1993}]%
        {blelloch1993nesl}
\bibfield{author}{\bibinfo{person}{Guy~E. Blelloch}.}
  \bibinfo{year}{1993}\natexlab{}.
\newblock \bibinfo{booktitle}{\emph{NESL: A Nested Data-Parallel Language
  (Version 2.6)}}.
\newblock \bibinfo{type}{{T}echnical {R}eport}. \bibinfo{address}{USA}.
\newblock


\bibitem[\protect\citeauthoryear{Bondhugula, Hartono, Ramanujam, and
  Sadayappan}{Bondhugula et~al\mbox{.}}{2008}]%
        {pluto}
\bibfield{author}{\bibinfo{person}{Uday Bondhugula}, \bibinfo{person}{Albert
  Hartono}, \bibinfo{person}{J. Ramanujam}, {and} \bibinfo{person}{P.
  Sadayappan}.} \bibinfo{year}{2008}\natexlab{}.
\newblock \showarticletitle{A Practical Automatic Polyhedral Parallelizer and
  Locality Optimizer}. In \bibinfo{booktitle}{\emph{Proceedings of the 29th ACM
  SIGPLAN Conference on Programming Language Design and Implementation}}
  \emph{(\bibinfo{series}{PLDI '08})}. \bibinfo{publisher}{Association for
  Computing Machinery}, \bibinfo{address}{New York, NY, USA},
  \bibinfo{pages}{101–113}.
\newblock
\showISBNx{9781595938602}
\urldef\tempurl%
\url{https://doi.org/10.1145/1375581.1375595}
\showDOI{\tempurl}


\bibitem[\protect\citeauthoryear{Brachth\"{a}user, Schuster, and
  Ostermann}{Brachth\"{a}user et~al\mbox{.}}{2020}]%
        {brachthauser2020effects}
\bibfield{author}{\bibinfo{person}{Jonathan~Immanuel Brachth\"{a}user},
  \bibinfo{person}{Philipp Schuster}, {and} \bibinfo{person}{Klaus Ostermann}.}
  \bibinfo{year}{2020}\natexlab{}.
\newblock \showarticletitle{Effects as Capabilities: Effect Handlers and
  Lightweight Effect Polymorphism}.
\newblock \bibinfo{journal}{\emph{Proc. ACM Program. Lang.}}
  \bibinfo{volume}{4}, \bibinfo{number}{OOPSLA}, Article
  \bibinfo{articleno}{126} (\bibinfo{date}{Nov.} \bibinfo{year}{2020}),
  \bibinfo{numpages}{30}~pages.
\newblock
\urldef\tempurl%
\url{https://doi.org/10.1145/3428194}
\showDOI{\tempurl}


\bibitem[\protect\citeauthoryear{Bradbury, Frostig, Hawkins, Johnson, Leary,
  Maclaurin, Necula, Paszke, Vander{P}las, Wanderman-{M}ilne, and
  Zhang}{Bradbury et~al\mbox{.}}{2018}]%
        {jax2018github}
\bibfield{author}{\bibinfo{person}{James Bradbury}, \bibinfo{person}{Roy
  Frostig}, \bibinfo{person}{Peter Hawkins}, \bibinfo{person}{Matthew~James
  Johnson}, \bibinfo{person}{Chris Leary}, \bibinfo{person}{Dougal Maclaurin},
  \bibinfo{person}{George Necula}, \bibinfo{person}{Adam Paszke},
  \bibinfo{person}{Jake Vander{P}las}, \bibinfo{person}{Skye
  Wanderman-{M}ilne}, {and} \bibinfo{person}{Qiao Zhang}.}
  \bibinfo{year}{2018}\natexlab{}.
\newblock \bibinfo{booktitle}{\emph{{JAX}: composable transformations of
  {P}ython+{N}um{P}y programs}}.
\newblock
\urldef\tempurl%
\url{http://github.com/google/jax}
\showURL{%
\tempurl}


\bibitem[\protect\citeauthoryear{Chakravarty, Keller, Lee, McDonell, and
  Grover}{Chakravarty et~al\mbox{.}}{2011}]%
        {chakravarty2011accelerate}
\bibfield{author}{\bibinfo{person}{Manuel M~T Chakravarty},
  \bibinfo{person}{Gabriele Keller}, \bibinfo{person}{Sean Lee},
  \bibinfo{person}{Trevor~L. McDonell}, {and} \bibinfo{person}{Vinod Grover}.}
  \bibinfo{year}{2011}\natexlab{}.
\newblock \showarticletitle{{Accelerating Haskell array codes with multicore
  GPUs}}. In \bibinfo{booktitle}{\emph{DAMP '11: The 6th workshop on
  Declarative Aspects of Multicore Programming}}. \bibinfo{publisher}{ACM}.
\newblock


\bibitem[\protect\citeauthoryear{Che, Boyer, Meng, Tarjan, Sheaffer, Lee, and
  Skadron}{Che et~al\mbox{.}}{2009}]%
        {che2009rodinia}
\bibfield{author}{\bibinfo{person}{Shuai Che}, \bibinfo{person}{Michael Boyer},
  \bibinfo{person}{Jiayuan Meng}, \bibinfo{person}{David Tarjan},
  \bibinfo{person}{Jeremy~W Sheaffer}, \bibinfo{person}{Sang-Ha Lee}, {and}
  \bibinfo{person}{Kevin Skadron}.} \bibinfo{year}{2009}\natexlab{}.
\newblock \showarticletitle{Rodinia: A benchmark suite for heterogeneous
  computing}. In \bibinfo{booktitle}{\emph{2009 IEEE international symposium on
  workload characterization (IISWC)}}. IEEE, \bibinfo{pages}{44--54}.
\newblock


\bibitem[\protect\citeauthoryear{Elliott}{Elliott}{2018}]%
        {elliott2018ad}
\bibfield{author}{\bibinfo{person}{Conal Elliott}.}
  \bibinfo{year}{2018}\natexlab{}.
\newblock \showarticletitle{The Simple Essence of Automatic Differentiation}.
\newblock \bibinfo{journal}{\emph{Proc. ACM Program. Lang.}}
  \bibinfo{volume}{2}, \bibinfo{number}{ICFP}, Article \bibinfo{articleno}{70}
  (\bibinfo{date}{July} \bibinfo{year}{2018}), \bibinfo{numpages}{29}~pages.
\newblock
\urldef\tempurl%
\url{https://doi.org/10.1145/3236765}
\showDOI{\tempurl}


\bibitem[\protect\citeauthoryear{Frostig, Johnson, Maclaurin, Paszke, and
  Radul}{Frostig et~al\mbox{.}}{2021}]%
        {frostig2021decomposing}
\bibfield{author}{\bibinfo{person}{Roy Frostig}, \bibinfo{person}{Matthew
  Johnson}, \bibinfo{person}{Dougal Maclaurin}, \bibinfo{person}{Adam Paszke},
  {and} \bibinfo{person}{Alexey Radul}.} \bibinfo{year}{2021}\natexlab{}.
\newblock \showarticletitle{Decomposing reverse-mode automatic
  differentiation}. In \bibinfo{booktitle}{\emph{LAFI '21: POPL 2021 workshop
  on Languages for Inference}}.
\newblock


\bibitem[\protect\citeauthoryear{Griewank and Walther}{Griewank and
  Walther}{2008}]%
        {griewank2008evaluating}
\bibfield{author}{\bibinfo{person}{Andreas Griewank} {and}
  \bibinfo{person}{Andrea Walther}.} \bibinfo{year}{2008}\natexlab{}.
\newblock \bibinfo{booktitle}{\emph{Evaluating Derivatives: Principles and
  Techniques of Algorithmic Differentiation} (\bibinfo{edition}{second} ed.)}.
\newblock \bibinfo{publisher}{Society for Industrial and Applied Mathematics},
  \bibinfo{address}{USA}.
\newblock
\showISBNx{0898716594}


\bibitem[\protect\citeauthoryear{Grosser, Gr{\"o}{\ss}linger, and
  Lengauer}{Grosser et~al\mbox{.}}{2012}]%
        {grosser2012polly}
\bibfield{author}{\bibinfo{person}{Tobias Grosser}, \bibinfo{person}{Armin
  Gr{\"o}{\ss}linger}, {and} \bibinfo{person}{C. Lengauer}.}
  \bibinfo{year}{2012}\natexlab{}.
\newblock \showarticletitle{Polly - Performing Polyhedral Optimizations on a
  Low-Level Intermediate Representation}.
\newblock \bibinfo{journal}{\emph{Parallel Process. Lett.}}
  \bibinfo{volume}{22} (\bibinfo{year}{2012}).
\newblock


\bibitem[\protect\citeauthoryear{Harris, Millman, van~der Walt, Gommers,
  Virtanen, Cournapeau, Wieser, Taylor, Berg, Smith, Kern, Picus, Hoyer, van
  Kerkwijk, Brett, Haldane, del R{'{\i}}o, Wiebe, Peterson,
  G{'{e}}rard-Marchant, Sheppard, Reddy, Weckesser, Abbasi, Gohlke, and
  Oliphant}{Harris et~al\mbox{.}}{2020}]%
        {harris2020numpy}
\bibfield{author}{\bibinfo{person}{Charles~R. Harris},
  \bibinfo{person}{K.~Jarrod Millman}, \bibinfo{person}{St{'{e}}fan~J. van~der
  Walt}, \bibinfo{person}{Ralf Gommers}, \bibinfo{person}{Pauli Virtanen},
  \bibinfo{person}{David Cournapeau}, \bibinfo{person}{Eric Wieser},
  \bibinfo{person}{Julian Taylor}, \bibinfo{person}{Sebastian Berg},
  \bibinfo{person}{Nathaniel~J. Smith}, \bibinfo{person}{Robert Kern},
  \bibinfo{person}{Matti Picus}, \bibinfo{person}{Stephan Hoyer},
  \bibinfo{person}{Marten~H. van Kerkwijk}, \bibinfo{person}{Matthew Brett},
  \bibinfo{person}{Allan Haldane}, \bibinfo{person}{Jaime~Fern{'{a}}ndez del
  R{'{\i}}o}, \bibinfo{person}{Mark Wiebe}, \bibinfo{person}{Pearu Peterson},
  \bibinfo{person}{Pierre G{'{e}}rard-Marchant}, \bibinfo{person}{Kevin
  Sheppard}, \bibinfo{person}{Tyler Reddy}, \bibinfo{person}{Warren Weckesser},
  \bibinfo{person}{Hameer Abbasi}, \bibinfo{person}{Christoph Gohlke}, {and}
  \bibinfo{person}{Travis~E. Oliphant}.} \bibinfo{year}{2020}\natexlab{}.
\newblock \showarticletitle{Array programming with {NumPy}}.
\newblock \bibinfo{journal}{\emph{Nature}} \bibinfo{volume}{585},
  \bibinfo{number}{7825} (\bibinfo{date}{Sept.} \bibinfo{year}{2020}),
  \bibinfo{pages}{357--362}.
\newblock
\urldef\tempurl%
\url{https://doi.org/10.1038/s41586-020-2649-2}
\showDOI{\tempurl}


\bibitem[\protect\citeauthoryear{Hascoet and Pascual}{Hascoet and
  Pascual}{2013}]%
        {hascoet2013tapenade}
\bibfield{author}{\bibinfo{person}{Laurent Hascoet} {and}
  \bibinfo{person}{Val{\'e}rie Pascual}.} \bibinfo{year}{2013}\natexlab{}.
\newblock \showarticletitle{The {T}apenade automatic differentiation tool:
  principles, model, and specification}.
\newblock \bibinfo{journal}{\emph{ACM Transactions on Mathematical Software
  (TOMS)}} \bibinfo{volume}{39}, \bibinfo{number}{3} (\bibinfo{year}{2013}),
  \bibinfo{pages}{1--43}.
\newblock


\bibitem[\protect\citeauthoryear{Henriksen, Hellfritzsch, Sadayappan, and
  Oancea}{Henriksen et~al\mbox{.}}{2020}]%
        {henriksen2020histograms}
\bibfield{author}{\bibinfo{person}{Troels Henriksen}, \bibinfo{person}{Sune
  Hellfritzsch}, \bibinfo{person}{Ponnuswamy Sadayappan}, {and}
  \bibinfo{person}{Cosmin Oancea}.} \bibinfo{year}{2020}\natexlab{}.
\newblock \showarticletitle{Compiling Generalized Histograms for GPU}. In
  \bibinfo{booktitle}{\emph{Proceedings of the International Conference for
  High Performance Computing, Networking, Storage and Analysis}}
  \emph{(\bibinfo{series}{SC '20})}. \bibinfo{publisher}{IEEE Press}, Article
  \bibinfo{articleno}{97}, \bibinfo{numpages}{14}~pages.
\newblock
\showISBNx{9781728199986}


\bibitem[\protect\citeauthoryear{Henriksen, Larsen, and Oancea}{Henriksen
  et~al\mbox{.}}{2016}]%
        {henriksen2016design}
\bibfield{author}{\bibinfo{person}{Troels Henriksen},
  \bibinfo{person}{Ken~Friis Larsen}, {and} \bibinfo{person}{Cosmin~E.
  Oancea}.} \bibinfo{year}{2016}\natexlab{}.
\newblock \showarticletitle{Design and GPGPU Performance of Futhark's Redomap
  Construct}. In \bibinfo{booktitle}{\emph{Proceedings of the 3rd ACM SIGPLAN
  International Workshop on Libraries, Languages, and Compilers for Array
  Programming}} \emph{(\bibinfo{series}{ARRAY 2016})}.
  \bibinfo{publisher}{Association for Computing Machinery},
  \bibinfo{address}{New York, NY, USA}, \bibinfo{pages}{17–24}.
\newblock
\showISBNx{9781450343848}
\urldef\tempurl%
\url{https://doi.org/10.1145/2935323.2935326}
\showDOI{\tempurl}


\bibitem[\protect\citeauthoryear{Henriksen, Serup, Elsman, Henglein, and
  Oancea}{Henriksen et~al\mbox{.}}{2017}]%
        {henriksen2017futhark}
\bibfield{author}{\bibinfo{person}{Troels Henriksen}, \bibinfo{person}{Niels~GW
  Serup}, \bibinfo{person}{Martin Elsman}, \bibinfo{person}{Fritz Henglein},
  {and} \bibinfo{person}{Cosmin~E Oancea}.} \bibinfo{year}{2017}\natexlab{}.
\newblock \showarticletitle{Futhark: purely functional GPU-programming with
  nested parallelism and in-place array updates}. In
  \bibinfo{booktitle}{\emph{Proceedings of the 38th ACM SIGPLAN Conference on
  Programming Language Design and Implementation}}. \bibinfo{pages}{556--571}.
\newblock


\bibitem[\protect\citeauthoryear{Hovgaard, Henriksen, and Elsman}{Hovgaard
  et~al\mbox{.}}{2018}]%
        {hovgaard2018high}
\bibfield{author}{\bibinfo{person}{Anders~Kiel Hovgaard},
  \bibinfo{person}{Troels Henriksen}, {and} \bibinfo{person}{Martin Elsman}.}
  \bibinfo{year}{2018}\natexlab{}.
\newblock \showarticletitle{High-Performance Defunctionalisation in Futhark}.
  In \bibinfo{booktitle}{\emph{International Symposium on Trends in Functional
  Programming}}. Springer, \bibinfo{pages}{136--156}.
\newblock


\bibitem[\protect\citeauthoryear{Hu, Anderson, Li, Sun, Carr, Ragan-Kelley, and
  Durand}{Hu et~al\mbox{.}}{2020}]%
        {hu2020difftaichi}
\bibfield{author}{\bibinfo{person}{Yuanming Hu}, \bibinfo{person}{Luke
  Anderson}, \bibinfo{person}{Tzu-Mao Li}, \bibinfo{person}{Qi Sun},
  \bibinfo{person}{Nathan Carr}, \bibinfo{person}{Jonathan Ragan-Kelley}, {and}
  \bibinfo{person}{Fredo Durand}.} \bibinfo{year}{2020}\natexlab{}.
\newblock \showarticletitle{DiffTaichi: Differentiable Programming for Physical
  Simulation}. In \bibinfo{booktitle}{\emph{International Conference on
  Learning Representations}}.
\newblock
\urldef\tempurl%
\url{https://openreview.net/forum?id=B1eB5xSFvr}
\showURL{%
\tempurl}


\bibitem[\protect\citeauthoryear{Hu, Li, Anderson, Ragan-Kelley, and Durand}{Hu
  et~al\mbox{.}}{2019}]%
        {hu2019taichi}
\bibfield{author}{\bibinfo{person}{Yuanming Hu}, \bibinfo{person}{Tzu-Mao Li},
  \bibinfo{person}{Luke Anderson}, \bibinfo{person}{Jonathan Ragan-Kelley},
  {and} \bibinfo{person}{Fr\'{e}do Durand}.} \bibinfo{year}{2019}\natexlab{}.
\newblock \showarticletitle{Taichi: A Language for High-Performance Computation
  on Spatially Sparse Data Structures}.
\newblock \bibinfo{journal}{\emph{ACM Trans. Graph.}} \bibinfo{volume}{38},
  \bibinfo{number}{6}, Article \bibinfo{articleno}{201} (\bibinfo{date}{Nov.}
  \bibinfo{year}{2019}), \bibinfo{numpages}{16}~pages.
\newblock
\showISSN{0730-0301}
\urldef\tempurl%
\url{https://doi.org/10.1145/3355089.3356506}
\showDOI{\tempurl}


\bibitem[\protect\citeauthoryear{H{\"u}ckelheim, Kukreja, Narayanan, Luporini,
  Gorman, and Hovland}{H{\"u}ckelheim et~al\mbox{.}}{2019}]%
        {huckelheim2019automatic}
\bibfield{author}{\bibinfo{person}{Jan H{\"u}ckelheim}, \bibinfo{person}{Navjot
  Kukreja}, \bibinfo{person}{Sri Hari~Krishna Narayanan},
  \bibinfo{person}{Fabio Luporini}, \bibinfo{person}{Gerard Gorman}, {and}
  \bibinfo{person}{Paul Hovland}.} \bibinfo{year}{2019}\natexlab{}.
\newblock \showarticletitle{Automatic differentiation for adjoint stencil
  loops}. In \bibinfo{booktitle}{\emph{Proceedings of the 48th International
  Conference on Parallel Processing}}. \bibinfo{pages}{1--10}.
\newblock


\bibitem[\protect\citeauthoryear{Innes}{Innes}{2018}]%
        {Zygote.jl-2018}
\bibfield{author}{\bibinfo{person}{Michael Innes}.}
  \bibinfo{year}{2018}\natexlab{}.
\newblock \showarticletitle{Don't Unroll Adjoint: Differentiating SSA-Form
  Programs}.
\newblock \bibinfo{journal}{\emph{CoRR}}  \bibinfo{volume}{abs/1810.07951}
  (\bibinfo{year}{2018}).
\newblock
\showeprint[arxiv]{1810.07951}
\urldef\tempurl%
\url{http://arxiv.org/abs/1810.07951}
\showURL{%
\tempurl}


\bibitem[\protect\citeauthoryear{Iverson}{Iverson}{1962}]%
        {iverson1962apl}
\bibfield{author}{\bibinfo{person}{Kenneth~E. Iverson}.}
  \bibinfo{year}{1962}\natexlab{}.
\newblock \bibinfo{booktitle}{\emph{A Programming Language}}.
\newblock \bibinfo{publisher}{John Wiley \& Sons, Inc.},
  \bibinfo{address}{USA}.
\newblock
\showISBNx{0471430145}


\bibitem[\protect\citeauthoryear{Larsen and Henriksen}{Larsen and
  Henriksen}{2017}]%
        {larsen2017strategies}
\bibfield{author}{\bibinfo{person}{Rasmus~Wriedt Larsen} {and}
  \bibinfo{person}{Troels Henriksen}.} \bibinfo{year}{2017}\natexlab{}.
\newblock \showarticletitle{Strategies for Regular Segmented Reductions on
  GPU}. In \bibinfo{booktitle}{\emph{Proceedings of the 6th ACM SIGPLAN
  International Workshop on Functional High-Performance Computing}}
  \emph{(\bibinfo{series}{FHPC 2017})}. \bibinfo{publisher}{Association for
  Computing Machinery}, \bibinfo{address}{New York, NY, USA},
  \bibinfo{pages}{42–52}.
\newblock
\showISBNx{9781450351812}
\urldef\tempurl%
\url{https://doi.org/10.1145/3122948.3122952}
\showDOI{\tempurl}


\bibitem[\protect\citeauthoryear{Launchbury and Peyton~Jones}{Launchbury and
  Peyton~Jones}{1994}]%
        {launchbury1994st}
\bibfield{author}{\bibinfo{person}{John Launchbury} {and}
  \bibinfo{person}{Simon~L. Peyton~Jones}.} \bibinfo{year}{1994}\natexlab{}.
\newblock \showarticletitle{Lazy Functional State Threads}. In
  \bibinfo{booktitle}{\emph{Proceedings of the ACM SIGPLAN 1994 Conference on
  Programming Language Design and Implementation}} \emph{(\bibinfo{series}{PLDI
  '94})}. \bibinfo{publisher}{Association for Computing Machinery},
  \bibinfo{address}{New York, NY, USA}, \bibinfo{pages}{24–35}.
\newblock
\showISBNx{089791662X}
\urldef\tempurl%
\url{https://doi.org/10.1145/178243.178246}
\showDOI{\tempurl}


\bibitem[\protect\citeauthoryear{Leijen}{Leijen}{2014}]%
        {leijen2014koka}
\bibfield{author}{\bibinfo{person}{Daan Leijen}.}
  \bibinfo{year}{2014}\natexlab{}.
\newblock \showarticletitle{Koka: Programming with Row Polymorphic Effect
  Types}.
\newblock \bibinfo{journal}{\emph{Electronic Proceedings in Theoretical
  Computer Science}}  \bibinfo{volume}{153} (\bibinfo{date}{Jun}
  \bibinfo{year}{2014}), \bibinfo{pages}{100–126}.
\newblock
\showISSN{2075-2180}
\urldef\tempurl%
\url{https://doi.org/10.4204/eptcs.153.8}
\showDOI{\tempurl}


\bibitem[\protect\citeauthoryear{Li, Gharbi, Adams, Durand, and
  Ragan-Kelley}{Li et~al\mbox{.}}{2018}]%
        {li2018halidead}
\bibfield{author}{\bibinfo{person}{Tzu-Mao Li}, \bibinfo{person}{Micha{\"e}l
  Gharbi}, \bibinfo{person}{Andrew Adams}, \bibinfo{person}{Fr{\'e}do Durand},
  {and} \bibinfo{person}{Jonathan Ragan-Kelley}.}
  \bibinfo{year}{2018}\natexlab{}.
\newblock \showarticletitle{Differentiable programming for image processing and
  deep learning in {Halide}}.
\newblock \bibinfo{journal}{\emph{ACM Trans. Graph. (Proc. SIGGRAPH)}}
  \bibinfo{volume}{37}, \bibinfo{number}{4} (\bibinfo{year}{2018}),
  \bibinfo{pages}{139:1--139:13}.
\newblock


\bibitem[\protect\citeauthoryear{Maclaurin, Duvenaud, and Adams}{Maclaurin
  et~al\mbox{.}}{2014}]%
        {maclaurin2015autograd}
\bibfield{author}{\bibinfo{person}{Dougal Maclaurin}, \bibinfo{person}{David
  Duvenaud}, {and} \bibinfo{person}{Ryan~P Adams}.}
  \bibinfo{year}{2014}\natexlab{}.
\newblock \showarticletitle{Autograd: Effortless gradients in numpy}
  \emph{(\bibinfo{series}{ICML '15 AutoML workshop})}. \bibinfo{pages}{5}.
\newblock


\bibitem[\protect\citeauthoryear{Manzyuk, Pearlmutter, Radul, Rush, and
  Siskind}{Manzyuk et~al\mbox{.}}{2019}]%
        {manzyuk2019perturbationconfusion}
\bibfield{author}{\bibinfo{person}{Oleksandr Manzyuk},
  \bibinfo{person}{Barak~A. Pearlmutter}, \bibinfo{person}{Alexey~Andreyevich
  Radul}, \bibinfo{person}{David~R. Rush}, {and} \bibinfo{person}{Jeffrey~Mark
  Siskind}.} \bibinfo{year}{2019}\natexlab{}.
\newblock \showarticletitle{Perturbation confusion in forward automatic
  differentiation of higher-order functions}.
\newblock \bibinfo{journal}{\emph{Journal of Functional Programming}}
  \bibinfo{volume}{29} (\bibinfo{year}{2019}), \bibinfo{pages}{e12}.
\newblock
\urldef\tempurl%
\url{https://doi.org/10.1017/S095679681900008X}
\showDOI{\tempurl}


\bibitem[\protect\citeauthoryear{McDonell, Chakravarty, Keller, and
  Lippmeier}{McDonell et~al\mbox{.}}{2013}]%
        {mcdonell2013optimizing}
\bibfield{author}{\bibinfo{person}{Trevor~L. McDonell}, \bibinfo{person}{Manuel
  M~T Chakravarty}, \bibinfo{person}{Gabriele Keller}, {and}
  \bibinfo{person}{Ben Lippmeier}.} \bibinfo{year}{2013}\natexlab{}.
\newblock \showarticletitle{{Optimising Purely Functional GPU Programs}}. In
  \bibinfo{booktitle}{\emph{ICFP '13: The 18th ACM SIGPLAN International
  Conference on Functional Programming}}. \bibinfo{publisher}{ACM}.
\newblock


\bibitem[\protect\citeauthoryear{Milner, Tofte, and Macqueen}{Milner
  et~al\mbox{.}}{1997}]%
        {milner1997ml}
\bibfield{author}{\bibinfo{person}{Robin Milner}, \bibinfo{person}{Mads Tofte},
  {and} \bibinfo{person}{David Macqueen}.} \bibinfo{year}{1997}\natexlab{}.
\newblock \bibinfo{booktitle}{\emph{The Definition of Standard ML}}.
\newblock \bibinfo{publisher}{MIT Press}, \bibinfo{address}{Cambridge, MA,
  USA}.
\newblock
\showISBNx{0262631814}


\bibitem[\protect\citeauthoryear{Mitchell}{Mitchell}{2010}]%
        {mitchell2010rethinking}
\bibfield{author}{\bibinfo{person}{Neil Mitchell}.}
  \bibinfo{year}{2010}\natexlab{}.
\newblock \showarticletitle{Rethinking Supercompilation}. In
  \bibinfo{booktitle}{\emph{Proceedings of the 15th ACM SIGPLAN International
  Conference on Functional Programming}} \emph{(\bibinfo{series}{ICFP '10})}.
  \bibinfo{publisher}{Association for Computing Machinery},
  \bibinfo{address}{New York, NY, USA}, \bibinfo{pages}{309–320}.
\newblock
\showISBNx{9781605587943}
\urldef\tempurl%
\url{https://doi.org/10.1145/1863543.1863588}
\showDOI{\tempurl}


\bibitem[\protect\citeauthoryear{Najd, Lindley, Svenningsson, and Wadler}{Najd
  et~al\mbox{.}}{2016}]%
        {najd+:everything-old}
\bibfield{author}{\bibinfo{person}{Shayan Najd}, \bibinfo{person}{Sam Lindley},
  \bibinfo{person}{Josef Svenningsson}, {and} \bibinfo{person}{Philip Wadler}.}
  \bibinfo{year}{2016}\natexlab{}.
\newblock \showarticletitle{Everything Old is New Again: Quoted Domain-Specific
  Languages}. In \bibinfo{booktitle}{\emph{Proceedings of the 2016 ACM SIGPLAN
  Workshop on Partial Evaluation and Program Manipulation}}
  \emph{(\bibinfo{series}{PEPM '16})}. \bibinfo{publisher}{Association for
  Computing Machinery}, \bibinfo{address}{New York, NY, USA},
  \bibinfo{pages}{25–36}.
\newblock
\showISBNx{9781450340977}
\urldef\tempurl%
\url{https://doi.org/10.1145/2847538.2847541}
\showDOI{\tempurl}


\bibitem[\protect\citeauthoryear{Paszke, Gross, Massa, Lerer, Bradbury, Chanan,
  Killeen, Lin, Gimelshein, Antiga, Desmaison, K{\"{o}}pf, Yang, DeVito,
  Raison, Tejani, Chilamkurthy, Steiner, Fang, Bai, and Chintala}{Paszke
  et~al\mbox{.}}{2019}]%
        {paszke2019pytorch}
\bibfield{author}{\bibinfo{person}{Adam Paszke}, \bibinfo{person}{Sam Gross},
  \bibinfo{person}{Francisco Massa}, \bibinfo{person}{Adam Lerer},
  \bibinfo{person}{James Bradbury}, \bibinfo{person}{Gregory Chanan},
  \bibinfo{person}{Trevor Killeen}, \bibinfo{person}{Zeming Lin},
  \bibinfo{person}{Natalia Gimelshein}, \bibinfo{person}{Luca Antiga},
  \bibinfo{person}{Alban Desmaison}, \bibinfo{person}{Andreas K{\"{o}}pf},
  \bibinfo{person}{Edward Yang}, \bibinfo{person}{Zachary DeVito},
  \bibinfo{person}{Martin Raison}, \bibinfo{person}{Alykhan Tejani},
  \bibinfo{person}{Sasank Chilamkurthy}, \bibinfo{person}{Benoit Steiner},
  \bibinfo{person}{Lu Fang}, \bibinfo{person}{Junjie Bai}, {and}
  \bibinfo{person}{Soumith Chintala}.} \bibinfo{year}{2019}\natexlab{}.
\newblock \showarticletitle{PyTorch: An Imperative Style, High-Performance Deep
  Learning Library}. In \bibinfo{booktitle}{\emph{Advances in Neural
  Information Processing Systems 32: Annual Conference on Neural Information
  Processing Systems 2019, NeurIPS 2019, December 8-14, 2019, Vancouver, BC,
  Canada}}. \bibinfo{pages}{8024--8035}.
\newblock


\bibitem[\protect\citeauthoryear{Pearlmutter and Siskind}{Pearlmutter and
  Siskind}{2008}]%
        {pearlmutter2008lambda}
\bibfield{author}{\bibinfo{person}{Barak~A Pearlmutter} {and}
  \bibinfo{person}{Jeffrey~Mark Siskind}.} \bibinfo{year}{2008}\natexlab{}.
\newblock \showarticletitle{Reverse-mode AD in a functional framework: Lambda
  the ultimate backpropagator}.
\newblock \bibinfo{journal}{\emph{ACM Transactions on Programming Languages and
  Systems (TOPLAS)}} \bibinfo{volume}{30}, \bibinfo{number}{2}
  (\bibinfo{year}{2008}), \bibinfo{pages}{1--36}.
\newblock


\bibitem[\protect\citeauthoryear{Peyton~Jones}{Peyton~Jones}{2008}]%
        {peytonjones2008parallelhaskell}
\bibfield{author}{\bibinfo{person}{Simon Peyton~Jones}.}
  \bibinfo{year}{2008}\natexlab{}.
\newblock \showarticletitle{Harnessing the Multicores: Nested Data Parallelism
  in Haskell}. In \bibinfo{booktitle}{\emph{Proceedings of the 6th Asian
  Symposium on Programming Languages and Systems}}
  \emph{(\bibinfo{series}{APLAS '08})}. \bibinfo{publisher}{Springer-Verlag},
  \bibinfo{address}{Berlin, Heidelberg}, \bibinfo{pages}{138}.
\newblock
\showISBNx{9783540893295}
\urldef\tempurl%
\url{https://doi.org/10.1007/978-3-540-89330-1_10}
\showDOI{\tempurl}


\bibitem[\protect\citeauthoryear{Peyton~Jones and Marlow}{Peyton~Jones and
  Marlow}{2002}]%
        {peytonjones2002secrets}
\bibfield{author}{\bibinfo{person}{Simon Peyton~Jones} {and}
  \bibinfo{person}{Simon Marlow}.} \bibinfo{year}{2002}\natexlab{}.
\newblock \showarticletitle{Secrets of the Glasgow Haskell Compiler Inliner}.
\newblock \bibinfo{journal}{\emph{J. Funct. Program.}} \bibinfo{volume}{12},
  \bibinfo{number}{5} (\bibinfo{date}{July} \bibinfo{year}{2002}),
  \bibinfo{pages}{393–434}.
\newblock
\showISSN{0956-7968}
\urldef\tempurl%
\url{https://doi.org/10.1017/S0956796802004331}
\showDOI{\tempurl}


\bibitem[\protect\citeauthoryear{Peyton~Jones, Vytiniotis, Weirich, and
  Shields}{Peyton~Jones et~al\mbox{.}}{2007a}]%
        {peytonjones2007practical}
\bibfield{author}{\bibinfo{person}{Simon Peyton~Jones},
  \bibinfo{person}{Dimitrios Vytiniotis}, \bibinfo{person}{Stephanie Weirich},
  {and} \bibinfo{person}{Mark Shields}.} \bibinfo{year}{2007}\natexlab{a}.
\newblock \showarticletitle{Practical Type Inference for Arbitrary-Rank Types}.
\newblock \bibinfo{journal}{\emph{J. Funct. Program.}} \bibinfo{volume}{17},
  \bibinfo{number}{1} (\bibinfo{date}{Jan.} \bibinfo{year}{2007}),
  \bibinfo{pages}{1–82}.
\newblock
\showISSN{0956-7968}
\urldef\tempurl%
\url{https://doi.org/10.1017/S0956796806006034}
\showDOI{\tempurl}


\bibitem[\protect\citeauthoryear{Peyton~Jones, Vytiniotis, Weirich, and
  Shields}{Peyton~Jones et~al\mbox{.}}{2007b}]%
        {jones2007practicaltypeinference}
\bibfield{author}{\bibinfo{person}{Simon Peyton~Jones},
  \bibinfo{person}{Dimitrios Vytiniotis}, \bibinfo{person}{Stephanie Weirich},
  {and} \bibinfo{person}{Mark Shields}.} \bibinfo{year}{2007}\natexlab{b}.
\newblock \showarticletitle{Practical Type Inference for Arbitrary-Rank Types}.
\newblock \bibinfo{journal}{\emph{J. Funct. Program.}} \bibinfo{volume}{17},
  \bibinfo{number}{1} (\bibinfo{date}{Jan.} \bibinfo{year}{2007}),
  \bibinfo{pages}{1–82}.
\newblock
\showISSN{0956-7968}
\urldef\tempurl%
\url{https://doi.org/10.1017/S0956796806006034}
\showDOI{\tempurl}


\bibitem[\protect\citeauthoryear{Ragan-Kelley, Barnes, Adams, Paris, Durand,
  and Amarasinghe}{Ragan-Kelley et~al\mbox{.}}{2013}]%
        {ragankelley2013halide}
\bibfield{author}{\bibinfo{person}{Jonathan Ragan-Kelley},
  \bibinfo{person}{Connelly Barnes}, \bibinfo{person}{Andrew Adams},
  \bibinfo{person}{Sylvain Paris}, \bibinfo{person}{Fr\'{e}do Durand}, {and}
  \bibinfo{person}{Saman Amarasinghe}.} \bibinfo{year}{2013}\natexlab{}.
\newblock \showarticletitle{Halide: A Language and Compiler for Optimizing
  Parallelism, Locality, and Recomputation in Image Processing Pipelines}.
\newblock \bibinfo{journal}{\emph{SIGPLAN Not.}} \bibinfo{volume}{48},
  \bibinfo{number}{6} (\bibinfo{date}{June} \bibinfo{year}{2013}),
  \bibinfo{pages}{519–530}.
\newblock
\showISSN{0362-1340}
\urldef\tempurl%
\url{https://doi.org/10.1145/2499370.2462176}
\showDOI{\tempurl}


\bibitem[\protect\citeauthoryear{Ritchie and Sussman}{Ritchie and
  Sussman}{2021}]%
        {ritchie2021higherorderad}
\bibfield{author}{\bibinfo{person}{Sam Ritchie} {and}
  \bibinfo{person}{Gerald~Jay Sussman}.} \bibinfo{year}{2021}\natexlab{}.
\newblock \bibinfo{title}{AD on Higher Order Functions}.
\newblock
\newblock
\newblock
\shownote{Unpublished note.}


\bibitem[\protect\citeauthoryear{Roesch, Lyubomirsky, Weber, Pollock, Kirisame,
  Chen, and Tatlock}{Roesch et~al\mbox{.}}{2018}]%
        {roesch2018relay}
\bibfield{author}{\bibinfo{person}{Jared Roesch}, \bibinfo{person}{Steven
  Lyubomirsky}, \bibinfo{person}{Logan Weber}, \bibinfo{person}{Josh Pollock},
  \bibinfo{person}{Marisa Kirisame}, \bibinfo{person}{Tianqi Chen}, {and}
  \bibinfo{person}{Zachary Tatlock}.} \bibinfo{year}{2018}\natexlab{}.
\newblock \showarticletitle{Relay: A new {IR} for machine learning frameworks}.
  In \bibinfo{booktitle}{\emph{Proceedings of the 2nd ACM SIGPLAN International
  Workshop on Machine Learning and Programming Languages}}.
  \bibinfo{pages}{58--68}.
\newblock


\bibitem[\protect\citeauthoryear{Shaikhha, Fitzgibbon, Vytiniotis, and
  Peyton~Jones}{Shaikhha et~al\mbox{.}}{2019}]%
        {shaikhha2019fsmooth}
\bibfield{author}{\bibinfo{person}{Amir Shaikhha}, \bibinfo{person}{Andrew
  Fitzgibbon}, \bibinfo{person}{Dimitrios Vytiniotis}, {and}
  \bibinfo{person}{Simon Peyton~Jones}.} \bibinfo{year}{2019}\natexlab{}.
\newblock \showarticletitle{Efficient Differentiable Programming in a
  Functional Array-Processing Language}.
\newblock \bibinfo{journal}{\emph{Proc. ACM Program. Lang.}}
  \bibinfo{volume}{3}, \bibinfo{number}{ICFP}, Article \bibinfo{articleno}{97}
  (\bibinfo{date}{July} \bibinfo{year}{2019}), \bibinfo{numpages}{30}~pages.
\newblock
\urldef\tempurl%
\url{https://doi.org/10.1145/3341701}
\showDOI{\tempurl}


\bibitem[\protect\citeauthoryear{Slepak, Shivers, and Manolios}{Slepak
  et~al\mbox{.}}{2014}]%
        {slepak2014remora}
\bibfield{author}{\bibinfo{person}{Justin Slepak}, \bibinfo{person}{Olin
  Shivers}, {and} \bibinfo{person}{Panagiotis Manolios}.}
  \bibinfo{year}{2014}\natexlab{}.
\newblock \showarticletitle{An Array-Oriented Language with Static Rank
  Polymorphism}. In \bibinfo{booktitle}{\emph{Proceedings of the 23rd European
  Symposium on Programming Languages and Systems - Volume 8410}}.
  \bibinfo{publisher}{Springer-Verlag}, \bibinfo{address}{Berlin, Heidelberg},
  \bibinfo{pages}{27–46}.
\newblock
\showISBNx{9783642548321}
\urldef\tempurl%
\url{https://doi.org/10.1007/978-3-642-54833-8_3}
\showDOI{\tempurl}


\bibitem[\protect\citeauthoryear{Steuwer, Remmelg, and Dubach}{Steuwer
  et~al\mbox{.}}{2017}]%
        {steuwer2017lift}
\bibfield{author}{\bibinfo{person}{Michel Steuwer}, \bibinfo{person}{Toomas
  Remmelg}, {and} \bibinfo{person}{Christophe Dubach}.}
  \bibinfo{year}{2017}\natexlab{}.
\newblock \showarticletitle{Lift: A Functional Data-Parallel IR for
  High-Performance GPU Code Generation}. In
  \bibinfo{booktitle}{\emph{Proceedings of the 2017 International Symposium on
  Code Generation and Optimization}} \emph{(\bibinfo{series}{CGO '17})}.
  \bibinfo{publisher}{IEEE Press}, \bibinfo{pages}{74–85}.
\newblock
\showISBNx{9781509049318}


\bibitem[\protect\citeauthoryear{Stratton, Rodrigues, Sung, Obeid, Chang,
  Anssari, Liu, and Hwu}{Stratton et~al\mbox{.}}{2012}]%
        {stratton2012parboil}
\bibfield{author}{\bibinfo{person}{J.~A. Stratton},
  \bibinfo{person}{Christopher~I. Rodrigues}, \bibinfo{person}{I-Jui Sung},
  \bibinfo{person}{Nady Obeid}, \bibinfo{person}{Li-Wen Chang},
  \bibinfo{person}{N. Anssari}, \bibinfo{person}{G. Liu}, {and}
  \bibinfo{person}{W. Hwu}.} \bibinfo{year}{2012}\natexlab{}.
\newblock \showarticletitle{Parboil: A Revised Benchmark Suite for Scientific
  and Commercial Throughput Computing}.
\newblock


\bibitem[\protect\citeauthoryear{Swamy, Chen, Fournet, Strub, Bhargavan, and
  Yang}{Swamy et~al\mbox{.}}{2011}]%
        {swamy2011valuedependent}
\bibfield{author}{\bibinfo{person}{Nikhil Swamy}, \bibinfo{person}{Juan Chen},
  \bibinfo{person}{C\'{e}dric Fournet}, \bibinfo{person}{Pierre-Yves Strub},
  \bibinfo{person}{Karthikeyan Bhargavan}, {and} \bibinfo{person}{Jean Yang}.}
  \bibinfo{year}{2011}\natexlab{}.
\newblock \showarticletitle{Secure Distributed Programming with Value-Dependent
  Types}. In \bibinfo{booktitle}{\emph{Proceedings of the 16th ACM SIGPLAN
  International Conference on Functional Programming}}
  \emph{(\bibinfo{series}{ICFP '11})}. \bibinfo{publisher}{Association for
  Computing Machinery}, \bibinfo{address}{New York, NY, USA},
  \bibinfo{pages}{266–278}.
\newblock
\showISBNx{9781450308656}
\urldef\tempurl%
\url{https://doi.org/10.1145/2034773.2034811}
\showDOI{\tempurl}


\bibitem[\protect\citeauthoryear{Tokui, Okuta, Akiba, Niitani, Ogawa, Saito,
  Suzuki, Uenishi, Vogel, and Yamazaki~Vincent}{Tokui et~al\mbox{.}}{2019}]%
        {tokui2019chainer}
\bibfield{author}{\bibinfo{person}{Seiya Tokui}, \bibinfo{person}{Ryosuke
  Okuta}, \bibinfo{person}{Takuya Akiba}, \bibinfo{person}{Yusuke Niitani},
  \bibinfo{person}{Toru Ogawa}, \bibinfo{person}{Shunta Saito},
  \bibinfo{person}{Shuji Suzuki}, \bibinfo{person}{Kota Uenishi},
  \bibinfo{person}{Brian Vogel}, {and} \bibinfo{person}{Hiroyuki
  Yamazaki~Vincent}.} \bibinfo{year}{2019}\natexlab{}.
\newblock \showarticletitle{Chainer: A deep learning framework for accelerating
  the research cycle}. In \bibinfo{booktitle}{\emph{Proceedings of the 25th ACM
  SIGKDD International Conference on Knowledge Discovery \& Data Mining}}.
  \bibinfo{pages}{2002--2011}.
\newblock


\bibitem[\protect\citeauthoryear{Vasilache, Zinenko, Theodoridis, Goyal,
  DeVito, Moses, Verdoolaege, Adams, and Cohen}{Vasilache
  et~al\mbox{.}}{2018}]%
        {vasilache2018tensor}
\bibfield{author}{\bibinfo{person}{Nicolas Vasilache},
  \bibinfo{person}{Oleksandr Zinenko}, \bibinfo{person}{Theodoros Theodoridis},
  \bibinfo{person}{Priya Goyal}, \bibinfo{person}{Zachary DeVito},
  \bibinfo{person}{William~S. Moses}, \bibinfo{person}{Sven Verdoolaege},
  \bibinfo{person}{Andrew Adams}, {and} \bibinfo{person}{Albert Cohen}.}
  \bibinfo{year}{2018}\natexlab{}.
\newblock \bibinfo{title}{Tensor Comprehensions: Framework-Agnostic
  High-Performance Machine Learning Abstractions}.
\newblock
\newblock
\showeprint[arxiv]{cs.PL/1802.04730}


\end{thebibliography}

\ifextended
%% Appendix
\newpage
\appendix
\section{Full Rules}
\label{sec:full-rules}

Here we present complete rules for type checking (\Cref{fig:typerules-full}), type class checking (\Cref{fig:typerules-constraints}), and simplification (\Cref{fig:defunctionalization-full}) of the subset of Dex presented in the main text.

The argument that these simplification rules cover all cases (except as mentioned in the main text) is subtle, and relies on type-correctness and on typeclass constrants.
For example, a value of any type $\tau$ that satisfies $\ttt{Data}\ \tau$ is already simplified.

We do something similar with \ttt{case} expressions (rule \tsc{SCase}) as with \ttt{for}. The key difference is that when calculating the variables we need to bind we have to additionally
take into account the variable bound by the {\tt Left} or {\tt Right} pattern ($x$).
The remaining simplification rules concern themselves with direct simplifications, and with the fact that the action of a \ttt{runState} or \ttt{runAccum} should itself be simplified.
The latter two are somewhat involved because, even though the state is restricted to a type satisfying a \ttt{Data} or \ttt{VectorSpace} constraint and therefore not a function, the action is free to return a value of function type.

\begin{figure}[h]\footnotesize
\framebox{$\vdash_{\ttt{IdxSet}} \tau$}
\[
\Infer{}{}{\vdash_{\ttt{IdxSet}}~\ttt{Unit}}\qquad
\Infer{}{}{\vdash_{\ttt{IdxSet}}~\ttt{Fin}~v}\qquad
\Infer{}{\vdash_{\ttt{IdxSet}}~\tau_1 \qquad \vdash_{\ttt{IdxSet}}~\tau_2}{\vdash_{\ttt{IdxSet}}~(\tau_1,\tau_2)}\qquad
\Infer{}{\vdash_{\ttt{IdxSet}}~\tau_1 \qquad \vdash_{\ttt{IdxSet}}~\tau_2}{\vdash_{\ttt{IdxSet}}~\ttt{Either}~\tau_1~\tau_2}
\]

\vspace{1em}\framebox{$\vdash_{\ttt{Data}} \tau$}
\[
\Infer{}{}{\vdash_{\ttt{Data}}~\ttt{Unit}}\qquad
\Infer{}{}{\vdash_{\ttt{Data}}~\ttt{Int}}\qquad
\Infer{}{}{\vdash_{\ttt{Data}}~\ttt{Float}}\qquad
\Infer{}{}{\vdash_{\ttt{Data}}~\ttt{Fin}~v}\qquad
\]\[
\Infer{}{\vdash_{\ttt{Data}}~\tau_2}{\vdash_{\ttt{Data}}~ (\tau_1 \Rightarrow \tau_2)}\qquad
\Infer{}{\vdash_{\ttt{Data}}~\tau_1 \qquad \vdash_{\ttt{Data}}~\tau_2}{\vdash_{\ttt{Data}}~(\tau_1,\tau_2)}\qquad
\Infer{}{\vdash_{\ttt{Data}}~\tau_1 \qquad \vdash_{\ttt{Data}}~\tau_2}{\vdash_{\ttt{Data}}~\ttt{Either}~\tau_1~\tau_2}
\]

\vspace{1em}\framebox{$\vdash_{\ttt{VSpace}} \tau$}
\[
\Infer{}{}{\vdash_{\ttt{VSpace}}~\ttt{Float}}\qquad
\Infer{}{\vdash_{\ttt{VSpace}}~\tau_1 \qquad \vdash_{\ttt{VSpace}}~\tau_2}{\vdash_{\ttt{VSpace}}~(\tau_1,\tau_2)}\qquad
\Infer{}{\vdash_{\ttt{VSpace}}~\tau_2}{\vdash_{\ttt{VSpace}}~(\tau_1 \Rightarrow \tau_2)}
\]
\caption{Dex's rules for the \ttt{IndexSet}, \ttt{Data}, and \ttt{VectorSpace} constraints.}
\label{fig:typerules-constraints}
\end{figure}

\begin{figure}[h]\footnotesize
\renewcommand{\arraystretch}{3}
\begin{tabular}{cc}
\multicolumn{2}{c}{\framebox{$\Gamma \vdash v \ann\tau$}}\\
\renewcommand{\arraystretch}{1.6}
\begin{tabular}{c}
    $\Infer{TypeVar  }{}{x\ann\tau,~\Gamma \vdash x\ann\tau} $\\
    $\Infer{TypeType }{}{\Gamma\vdash \ttt{Type}  : \ttt{Type}} $\\
    $\Infer{TypeUnit }{}{\Gamma\vdash \ttt{Unit}  : \ttt{Type}} $\\
    $\Infer{TypeInt  }{}{\Gamma\vdash \ttt{Int}   : \ttt{Type}} $\\
    $\Infer{TypeFloat}{}{\Gamma\vdash \ttt{Float} : \ttt{Type}} $\\
    $\Infer{TypeFin  }{}
       {\Gamma\vdash \ttt{Fin} : \ttt{Int} \rightarrow \ttt{Type}} $\\
    $\Infer{TypeEither}{}
       {\Gamma\vdash \ttt{Either} : \ttt{Type} \rightarrow \ttt{Type} \rightarrow \ttt{Type}} $\\
\end{tabular}
&\renewcommand{\arraystretch}{3}
\begin{tabular}{c}
    $\Infer{TypeLam}
       {\epsilon,~x\ann\tau_1,~\Gamma \vdash e :  \subst{y}{x}{\tau_2}}
       {\Gamma \vdash (\lam x {\tau_1} e) : (y\ann\tau_1 \rightarrow \epsilon~\tau_2)} $\\
    $\Infer{TypeView}
       {     \ttt{Pure},~x\ann\tau_1,~\Gamma \vdash e :  \tau_2
       \quad \vdash_{\ttt{IdxSet}} \tau_1}
       {\Gamma \vdash (\view x {\tau_1} e) : (\tau_1 \Rightarrow \tau_2)} $\\
    $\Infer{TypePair}
       {\Gamma \vdash v_1 \ann \tau_1 \quad \Gamma \vdash v_2 \ann \tau_2}
       {\Gamma \vdash (v_1,~v_2) : (\tau_1 \times \tau_2)} $\\
    $\Infer{TypeLeft}
       {\Gamma \vdash \tau_2 : \ttt{Type} \quad \Gamma \vdash v \ann \tau_1}
       {\Gamma \vdash (\ttt{Left}~\tau_2~v) : \ttt{Either}~\tau_1~\tau_2} $\\
\end{tabular}
\\$\Infer{TypeFunction}
   {\Gamma \vdash \tau_1 : \ttt{Type} \quad
    x\ann\tau_1,~\Gamma \vdash \tau_2 : \ttt{Type}}
   {\Gamma \vdash (x\ann \tau_1 \rightarrow \epsilon~\tau_2) : \ttt{Type} }$
& $\Infer{TypeRight}
   {\Gamma \vdash \tau_1 : \ttt{Type} \quad \Gamma \vdash v \ann \tau_2}
   {\Gamma \vdash (\ttt{Right}~\tau_1~v) : \ttt{Either}~\tau_1~\tau_2} $\\
\multicolumn{2}{c}{\framebox{$\epsilon,~\Gamma \vdash e : \tau$}}
\\\Infer{TypeFst}
  {\Gamma\vdash v:(\tau_1 \times \tau_2)}
  {\epsilon,~\Gamma\vdash \ttt{fst}~v:\tau_1}
& \Infer{TypeSnd}
  {\Gamma\vdash v:(\tau_1 \times \tau_2)}
  {\epsilon,~\Gamma\vdash \ttt{snd}~v:\tau_2}
\\\Infer{TypeLet}
  {\epsilon,~\Gamma\vdash e_1 : \tau_1 \quad
   \epsilon,~x\ann\tau_1,~\Gamma\vdash e_2 : \tau_2 }
  {\epsilon,~\Gamma\vdash (\letx {x\ann\tau_1} {e_1} {e_2}) : \tau_2}
& \Infer{TypeApp}
  {\epsilon' \subseteq \epsilon \quad
   \Gamma\vdash v_1:(x\ann\tau_1 \rightarrow \epsilon'~\tau_2) \quad
   \Gamma\vdash v_2:\tau_1}
  {\epsilon,~\Gamma\vdash (v_1~v_2):\subst{x}{v_2}{\tau_2}}
\\\Infer{TypeIndex}
  {\Gamma\vdash v_1:(\tau_1 \Rightarrow \tau_2) \quad
   \Gamma\vdash v_2:\tau_1}
  {\epsilon,~\Gamma\vdash (v_1.v_2):\tau_2}
& \Infer{TypeSlice}
  {\Gamma\vdash v_1 : \ttt{Ref}~h~(\tau_1\Rightarrow\tau_2) \quad
   \Gamma\vdash v_2 : \tau_1}
  {\epsilon,~\Gamma\vdash (v_1 ! v_2 ) : \ttt{Ref}~h~\tau_2}
\\\Infer{TypeFor}
  {     \epsilon,~x\ann\tau_1,~\Gamma\vdash e : \tau_2
  \quad \vdash_{\ttt{IdxSet}} \tau_1}
  {\epsilon,~\Gamma\vdash (\forexpr x {\tau_1} e): (\tau_1\Rightarrow\tau_2)}
& \Infer{TypeAccumulate}
  {\Gamma \vdash v_1 : \ttt{Ref}~h~\tau \quad
   \Gamma \vdash v_2 : \tau }
  {\ttt{Accum}~h,~\epsilon,~\Gamma\vdash v_1~\ttt{+=}~v_2 : \ttt{Unit}}
\\\Infer{TypeGet}
  {\Gamma \vdash v : \ttt{Ref}~h~\tau }
  {\ttt{State}~h,~\epsilon,~\Gamma\vdash \ttt{get}~v : \tau}
& \Infer{TypePut}
  {\Gamma \vdash v_1 : \ttt{Ref}~h~\tau \quad
   \Gamma \vdash v_2 : \tau }
  {\ttt{State}~h,~\epsilon,~\Gamma\vdash \ttt{put}~v_1~v_2 : \ttt{Unit}}
\\\multicolumn{2}{c}{
\Infer{TypeRunState}
  {\Gamma \vdash v_1 : \tau_1 \quad
   \Gamma \vdash v_2 : (\ttt{Ref}~h~\tau_1 \rightarrow (\ttt{State}~h,~\epsilon)~\tau_2) \quad
   \vdash_{\ttt{Data}} \tau_1}
  {\epsilon,~\Gamma\vdash \ttt{runState}~v_1~v_2 : (\tau_2 \times \tau_1) }
}
\\\multicolumn{2}{c}{
\Infer{TypeRunAccum}
  {\Gamma \vdash v : (\ttt{Ref}~h~\tau_1 \rightarrow (\ttt{Accum}~h,~\epsilon)~\tau_2) \quad
   \vdash_{\ttt{VSpace}} \tau_1}
  {\epsilon,~\Gamma\vdash \ttt{runAccum}~v : (\tau_2 \times \tau_1) }
}
\\\multicolumn{2}{c}{
\Infer{TypeCase}
  {\Gamma\vdash v:(\ttt{Either}\ \tau_1\ \tau_2)\quad
   \epsilon,~x:\tau_1,~\Gamma\vdash e_1:\tau\quad
   \epsilon,~x:\tau_2,~\Gamma\vdash e_2:\tau}
  {\epsilon,~\Gamma\vdash \case v {e_1} {e_2}:\tau}
}
\\\multicolumn{2}{c}{
\Infer{TypeLinearize}
  {\Gamma\vdash v_1:\tau_1 \rightarrow \ttt{Pure}~\tau_2  \quad
   \Gamma\vdash v_2:\tau_1 \quad
   \vdash_{\ttt{VSpace}} \tau_1 \quad
   \vdash_{\ttt{VSpace}} \tau_2
   }
  {\epsilon,~\Gamma\vdash \ttt{linearize}~v_1~v_2:(\tau_2 \times (\tau_1 \rightarrow \tau_2))}
}
\\\multicolumn{2}{c}{
\Infer{TypeTranspose}
  {\Gamma\vdash v_1:\tau_1 \rightarrow \ttt{Pure}~\tau_2  \quad
   \Gamma\vdash v_2:\tau_2 \quad
   \vdash_{\ttt{VSpace}} \tau_1 \quad
   \vdash_{\ttt{VSpace}} \tau_2
   }
  {\epsilon,~\Gamma\vdash \ttt{transpose}~v_1~v_2:\tau_1}
}
\end{tabular}
\caption{Dex's typing rules for values and expressions.  Types $\tau$ are permitted to be arbitrary values (including variables, but excluding expressions) of type \ttt{Type}.  Type equality is checked structurally. For the purpose of type checking we model effects as capabilities in the spirit of \citet{brachthauser2020effects}. See \Cref{fig:typerules-constraints} for rules defining constraints on datatypes.}
\label{fig:typerules-full}
\end{figure}

\begin{figure}\footnotesize
\renewcommand{\arraystretch}{3}
\begin{tabular}{c}
\framebox{$e \defunc E^d , v$} \\
 \Infer{SVal}{~~}{v \defunc \bullet, v} \qquad
  \Infer{SExpr}
  {e^d : \tau^d\qquad x\ \mathrm{fresh}}
  {e^d \defunc \letx{x:\tau^d}{e^d}\bullet , x } \quad\quad
  \Infer{SApp}
  {\subst{x}{v}{e} \defunc E^d, v'}
  {(\lam x \tau e)~v \defunc E^d, v'} \quad\quad \\
\Infer{SLet}
  {             e_1  \defunc E^d_1, v_1  \qquad
   \subst{x}{v_1}{e_2} \defunc E^d_2, v_2}
  {\letx {x:\tau} {e_1} {e_2} \defunc E^d_1\circ E^d_2, v_2} \quad\quad
\Infer{SView}
  {\subst{x}{v}{e} \defunc E^d, v'}
  {(\view x \tau e).v \defunc E^d, v'} \\
\Infer{SFor}
  {e \defunc E^d, v  \qquad  \binders{E^d}\vdash v \triangleright \overline{x}^{1..n}
  \qquad x \notin \freevars{x_1,\ldots,x_n}
  \qquad y~\mathrm{fresh}}
  {\forexpr x \tau e \defunc
        (\letx {y} {\forexpr x \tau E^d[(x_1,\ldots,x_n)]} \bullet) ,
        ( \view x \tau \letx {(x_1,\ldots,x_n)} {y.x} v )  } \\
\Infer{SFst}
  {v_1 \defunc E^d, v}
  {\ttt{fst}~(v_1,~v_2) \defunc E^d, v} \quad\quad
  \Infer{SSnd}
  {v_2 \defunc E^d, v}
  {\ttt{snd}~(v_1,~v_2) \defunc E^d, v} \\
\Infer{SCase}
  {   e_1 \defunc E^d_1, v_1' \qquad
            (x{:}\tau_1),\binders{E^d_1} \vdash v_1' \triangleright \overline{x}^{1..n} \\\\
      e_2 \defunc E^d_2, v_2' \qquad
            (x{:}\tau_2),\binders{E^d_2} \vdash v_2' \triangleright \overline{y}^{1..m} \qquad
   z\ \mathrm{fresh}}
  {\case v {e_1} {e_2} \defunc \\
  (\letx {z} {\case v {E^d_1[\ttt{Left}~\sigma_2~(x_1,\ldots,x_n)]} {E^d_2[\ttt{Right}~\sigma_1~(y_1,\ldots,y_m)]}} \bullet), \\
  (\casebranch{z}{{\tt Left}~(x_1,\ldots,x_n) \rightarrow v_1'}
                 {{\tt Right}~(y_1,\ldots,y_m) \rightarrow v_2'})}\\
\Infer{SCaseLeft}
  {\subst{x}{v}{e_1} \defunc E^d, v'}
  {\case {\ttt{Left}\ \tau\ v} {e_1} {e_2} \defunc E^d, v'} \\
\Infer{SCaseRight}
  {\subst{x}{v}{e_2} \defunc E^d, v'}
  {\case {\ttt{Right}\ \tau\ v} {e_1} {e_2} \defunc E^d, v'} \\
\Infer{SRunState}
  {e \defunc E^d, v' \qquad (x{:}\tau),\binders{E^d} \vdash v' \triangleright \overline{x}^{1..n}
%   E^d, v' \saverestore v_r, v^d \qquad
%   (v_r~r) \defunc E^d_r, v'' \qquad
   \qquad s~\mathrm{fresh}}
  {\ttt{runState}~v~(\lam {h~x} \tau e)\defunc \\
   (\letx {((x_1,\ldots,x_n), s)} {\ttt{runState}~v~(\lam {h~x} \tau {E^d[(x_1,\ldots,x_n)]})} \bullet), (v', s) } \\
\Infer{SRunAccum}
  {e \defunc E^d, v' \qquad (x{:}\tau),\binders{E^d} \vdash v' \triangleright \overline{x}^{1..n}
%   E^d, v' \saverestore v_r, v^d \qquad
%   (v_r~r) \defunc E^d_r, v'' \qquad
   \qquad s~\mathrm{fresh}}
  {\ttt{runAccum}~v~(\lam {h~x} \tau e)\defunc \\
   (\letx {((x_1,\ldots,x_n), s)} {\ttt{runAccum}~v~(\lam {h~x} \tau {E^d[(x_1,\ldots,x_n)]})} \bullet), (v', s) } \\
\Infer{SLinearize}
     {e\defunc E_1^d,~v_1  \quad
      \linreify{\Gamma}{x}{E_1^d[v_1]} e' \quad
      e' \defunc E_2^d,~v_2
         }
 {\ttt{linearize}~(\lam x \tau e)~v\defunc \subst{x}{v}(E_2^d,~v_2)}
 \\
\Infer{STranspose}
     {e\defunc E_1^d,~v_1  \quad
      \transrule{x\rightarrow r}{E_1^d[v_1]} {v_t} e' \quad
      \ttt{yieldAccum}~(\action {h} r {\ttt{Ref}~h~\tau} e') \defunc E_2^d,~v_2
      \quad r,h~~\mathrm{fresh}
         }
 {\ttt{transpose}~(\lam x \tau e)~{v_t}\defunc E_2^d,~v_2}
 \\
\end{tabular}
\begin{tabular}{c}
\framebox{$\overline{x{:}\tau} \vdash v \triangleright \overline{y}$} \\
\Infer{Empty}{~~}{\cdot \vdash v \triangleright \emptyset} \qquad
\Infer{Used}
    { \overline{x{:}\tau_x} \vdash v \triangleright \overline{y} \qquad
        x_1 \in \freevars{v} \\\\\ \freevars{\tau_1} \cap \overline{y} = \emptyset}
    {(x_1{:}\tau_1),\overline{x{:}\tau} \vdash v \triangleright x_1,\overline{y}} \qquad
\Infer{NotUsed}
    { \overline{x{:}\tau_x} \vdash v \triangleright \overline{y} \qquad
        x_1 \notin \freevars{v}}
    {(x_1{:}\tau_1),\overline{x{:}\tau} \vdash v \triangleright \overline{y}}
\end{tabular}
\caption{Simplification rules. Types $\tau_1$, $\tau_2$, $\sigma_1$, and $\sigma_2$ that appear in the \tsc{SCase} rule can be calculated. The functions $\mathcal{L}_x$ and $\mathcal{T}_{x\rightarrow r}$ implement the source transformations for linearization (\Cref{fig:linearize}) and transposition (\Cref{fig:transpose}), respectively. The environment $\Gamma$ supplied to $\mathcal{L}_x$ is the type environment of the \ttt{linearize} call being simplified.}
\label{fig:defunctionalization-full}
\end{figure}
\fi
\end{document}